\documentclass[10pt,journal,compsoc]{IEEEtran}

\usepackage{amssymb}
\usepackage{amsfonts}
\usepackage{amsmath}
\usepackage{cite}
\usepackage[dvips]{graphicx}
\usepackage{color}
\usepackage{algorithm}
\usepackage{algorithmic}
\usepackage{verbatim}
\usepackage{multirow}
\usepackage{caption}

\begin{document}

\newtheorem{definition}{\it Definition}
\newtheorem{theorem}{\bf Theorem}
\newtheorem{lemma}{\it Lemma}
\newtheorem{corollary}{\it Corollary}
\newtheorem{remark}{\it Remark}
\newtheorem{example}{\it Example}
\newtheorem{case}{\bf Case Study}
\newtheorem{assumption}{\it Assumption}
\newtheorem{property}{\it Property}
\newtheorem{proposition}{\it Proposition}

\newcommand{\hP}[1]{{\boldsymbol h}_{{#1}{\bullet}}}
\newcommand{\hS}[1]{{\boldsymbol h}_{{\bullet}{#1}}}

\newcommand{\ba}{\boldsymbol{a}}
\newcommand{\baq}{\overline{q}}
\newcommand{\bA}{\boldsymbol{A}}
\newcommand{\bb}{\boldsymbol{b}}
\newcommand{\bB}{\boldsymbol{B}}
\newcommand{\bc}{\boldsymbol{c}}
\newcommand{\bcC}{\boldsymbol{\cal C}}
\newcommand{\bcO}{\boldsymbol{\cal O}}
\newcommand{\bd}{\boldsymbol{d}}
\newcommand{\bh}{\boldsymbol{h}}
\newcommand{\bH}{\boldsymbol{H}}
\newcommand{\bl}{\boldsymbol{l}}
\newcommand{\bm}{\boldsymbol{m}}
\newcommand{\bn}{\boldsymbol{n}}
\newcommand{\bo}{\boldsymbol{o}}
\newcommand{\bO}{\boldsymbol{O}}
\newcommand{\bp}{\boldsymbol{p}}
\newcommand{\bq}{\boldsymbol{q}}
\newcommand{\br}{\boldsymbol{r}}
\newcommand{\bR}{\boldsymbol{R}}
\newcommand{\bs}{\boldsymbol{s}}
\newcommand{\bS}{\boldsymbol{S}}
\newcommand{\bT}{\boldsymbol{T}}
\newcommand{\bu}{\boldsymbol{u}}
\newcommand{\bv}{\boldsymbol{v}}
\newcommand{\bw}{\boldsymbol{w}}
\newcommand{\bX}{\boldsymbol{X}}
\newcommand{\bZ}{\boldsymbol{Z}}
\newcommand{\bzero}{\boldsymbol{0}}

\newcommand{\balpha}{\boldsymbol{\alpha}}
\newcommand{\bbeta}{\boldsymbol{\beta}}
\newcommand{\bdeta}{\boldsymbol{\eta}}
\newcommand{\btau}{\boldsymbol{\tau}}
\newcommand{\bOmega}{\boldsymbol{\Omega}}
\newcommand{\bTheta}{\boldsymbol{\Theta}}
\newcommand{\bLambda}{\boldsymbol{\Lambda}}
\newcommand{\bphi}{\boldsymbol{\phi}}
\newcommand{\bPhi}{\boldsymbol{\Phi}}
\newcommand{\brho}{\boldsymbol{\rho}}
\newcommand{\btheta}{\boldsymbol{\theta}}
\newcommand{\bvarpi}{\boldsymbol{\varpi}}
\newcommand{\bpi}{\boldsymbol{\pi}}
\newcommand{\bpsi}{\boldsymbol{\psi}}
\newcommand{\bxi}{\boldsymbol{\xi}}
\newcommand{\bzeta}{\boldsymbol{\zeta}}
\newcommand{\bx}{\boldsymbol{x}}
\newcommand{\by}{\boldsymbol{y}}

\newcommand{\cA}{{\cal A}}
\newcommand{\bcA}{\boldsymbol{\cal A}}
\newcommand{\cB}{{\cal B}}
\newcommand{\cC}{{\cal C}}
\newcommand{\cD}{{\cal D}}
\newcommand{\cE}{{\cal E}}
\newcommand{\cG}{{\cal G}}
\newcommand{\cH}{{\cal H}}
\newcommand{\bcH}{\boldsymbol {\cal H}}
\newcommand{\cI}{{\cal I}}
\newcommand{\cK}{{\cal K}}
\newcommand{\cL}{{\cal L}}
\newcommand{\cM}{{\cal M}}
\newcommand{\cO}{{\cal O}}
\newcommand{\cR}{{\cal R}}
\newcommand{\cS}{{\cal S}}
\newcommand{\dcS}{\ddot{{\cal S}}}
\newcommand{\ds}{\ddot{{s}}}
\newcommand{\cT}{{\cal T}}
\newcommand{\cU}{{\cal U}}
\newcommand{\cY}{{\cal Y}}
\newcommand{\wt}[1]{\widetilde{#1}}

\newcommand{\mA}{\mathbb{A}}
\newcommand{\mE}{\mathbb{E}}
\newcommand{\mG}{\mathbb{G}}
\newcommand{\mR}{\mathbb{R}}
\newcommand{\mS}{\mathbb{S}}
\newcommand{\mU}{\mathbb{U}}
\newcommand{\mV}{\mathbb{V}}
\newcommand{\mW}{\mathbb{W}}

\newcommand{\uq}{\underline{q}}
\newcommand{\ubq}{\underline{\boldsymbol q}}

\newcommand{\red}[1]{\textcolor[rgb]{0,0,0}{#1}}
\newcommand{\gre}[1]{\textcolor[rgb]{0,1,0}{#1}}
\newcommand{\blu}[1]{\textcolor[rgb]{0,0,0}{#1}}

\title{\LARGE AdaptiveFog: A Modelling and Optimization Framework for Fog Computing in Intelligent Transportation Systems}
\author{Yong~Xiao, \IEEEmembership{Senior~Member,~IEEE} and Marwan Krunz, \IEEEmembership{Fellow, IEEE} 

\thanks{An abridged version of this paper was presented at the IEEE SECON Conference, Boston, MA, June 2019\cite{XY2019AdaptiveFog}.

Y. Xiao is with the School of Electronic Information and Communications at the Huazhong University of Science and Technology, Wuhan, China (e-mail: yongxiao@hust.edu.cn). Y. Xiao is also with Pazhou Lab, Guangzhou, China.

M. Krunz is with the Department of Electrical and Computer Engineering, the University of Arizona, Tucson, AZ (e-mail: krunz@arizona.edu). 

}
}



\maketitle
\begin{abstract}
Fog computing has been advocated as an enabling technology for computationally intensive services in smart connected vehicles. Most existing works focus on analyzing the queueing and workload processing latencies associated with fog computing, ignoring the fact that wireless access latency can sometimes dominate the overall latency. This motivates the work in this paper, where we report on a five-month measurement study of the wireless access latency between connected vehicles and a fog/cloud computing system supported by commercially available LTE networks. We propose {\em AdaptiveFog}, a novel framework for autonomous and dynamic switching between different LTE networks that implement a fog/cloud infrastructure. AdaptiveFog's main objective is to maximize the {\em service confidence level}, defined as the probability that the latency of a given service type is below some threshold.  To quantify the performance gap between different LTE networks, we introduce a novel statistical distance metric, called weighted Kantorovich-Rubinstein (K-R) distance. Two scenarios based on finite- and infinite-horizon optimization of short-term and long-term confidence are investigated. For each scenario, a simple threshold policy based on weighted K-R distance is proposed and proved to maximize the latency confidence for smart vehicles. Extensive analysis and simulations are performed based on our latency measurements. Our results show that AdaptiveFog achieves around 30\% to 50\%  improvement in the confidence levels of fog and cloud latencies, respectively.

\end{abstract}

\vspace{-0.1in}


\section{Introduction}
\label{Section_Introduction}

Ultra-reliable low-latency communication (URLLC) and processing are critical for supporting newly emerging 
intelligent transportation services (ITS), such as congestion avoidance, accident prevention, active control intervention, autonomous driving, and intelligent driver assistance (e.g., route computation, searchable maps, etc.). Due to the limited energy, processing, and storage capacities of the in-vehicle computer, various industry consortiums and standardization bodies (e.g., \cite{Sabella2017MECVehicular,NGMN2018V2X}) have been promoting the use of  
%
high-performance cloud data centers (CDCs) for external data storage (e.g., high-definition maps) and processing for connected vehicles. The physical connectivity between vehicles and CDCs may span several wireless and wired links, each having its own traffic dynamics, medium access mechanisms, and connection intermittency. As a result, the end-to-end communication path may exhibit unacceptable latency and link disruptions. This motivates the need for 
better solutions that are more suitable for fast-response and highly reliable services. 
Fog computing has recently been introduced as a promising approach to offload (partially or fully) the computational load from CDCs to local fog nodes, hence 
reducing the end-to-end latency\cite{Chiang2017FogBook}. Supporting smart vehicular applications via 
fog computing has the potential to significantly reduce the communication latency and improve service reliability\cite{Premsankar2018MECVehicular,Asadi2018V2I}. 

Fog computing has 
also been advocated by mobile network operators (MNOs) as a way to create new business opportunities, increase revenues, and reduce capital expenditures (CAPEX). Major MNOs, including AT\&T, Verizon, and Deutsche Telekom, have announced plans to integrate fog computing into their network infrastructure to support emerging applications such as robotic manufacturing, autonomous cars, and augmented/virtual reality (AR/VR)\cite{XY2018TactileInternet}. LTE is readily available to support 
high-speed and low-latency wireless solutions on a global scale, and therefore is a good candidate to facilitate MNO-based
fog computing. Recent rollouts of 5G networks are primarily based on the non-standalone (NSA) mode, which relies on existing LTE core networks and base stations\cite{5GPPPArchitecture2019v3}. \blu{Therefore, analyzing and modeling the wireless access latency in existing LTE systems allow us to better understand  the fundamental issues that would affect the practical implementation of URLLC in LTE and 5G as well as their evolution towards B5G and 6G systems\cite{XY20206GSelfLearn}.}
According to 3GPP, the round-trip-time (RTT) for user equipments (UEs) connected via  LTE networks should be in the order of 10 ms in ideal conditions\cite{3GPPLTETR25913}. Such RTT value is 
negligible when compared to other types of latencies in a fog computing system, including data processing and queueing delays. Unfortunately, recent reports as well as our own measurements suggest that the 10 ms target latency is unattainable 
by most existing commercial  LTE operators. In fact, recent studies\cite{Hadzic2017MECePC, Mir2014LTEWiFiforVeh, Xu2017LTEVehicular} indicate that the wireless connection between moving a vehicle and the LTE  network can sometimes experience frequent disconnections, retransmission, and high wireless access latency that dominate the overall end-to-end latency.  

While there has been numerous studies of the wireless access latency in LTE networks (e.g., \cite{Hadzic2017MECePC}), to date there has been no 
long-term systematic study of the latency between moving vehicles and cloud/fog servers over commercial LTE networks. In fact, due to the 
time-varying network topology, diverse application requirements, and highly dynamic traffic demand, 
modeling and optimizing the latency in an LTE-based vehicular systems are quite difficult.

This paper first empirically analyzes the observed latency 
in vehicle-to-cloud/fog solutions for connected vehicular systems supported by a multi-operator LTE network.  
Based on this analysis, we propose a novel optimization framework, called {\em AdaptiveFog}, which can be used by a smart vehicle to dynamically switch between MNO networks that offer fog and cloud services on the move. 
We use a smart phone app, built on Android API and run on a Google Pixel 2 phone located inside 
a vehicle, to obtain five-month-long measurements of fog/cloud latencies associated with 
two major LTE MNOs. 
These measurements are used to evaluate the impact of handover, driving speed, MNO network, fog/cloud server, and location on the service latency.

Based on our measurements, we observe that none of the MNOs consistently offers better latency performance than the other. Also, the instantaneous latency varies significantly between measurements. However, the statistical features, including the empirical probability distribution function (PDF) of latency, remain relatively stationary at any given vehicle location.
%
Accordingly, we investigate the confidence level of a vehicular service, defined by the probability that a tolerable latency threshold can be guaranteed for the supported-type of service, across a city-wide geographical area. An empirical spatial statistical model is established using our dataset. We observe that although the difference between the mean values of cloud and fog latencies can be as low as 10-20 ms, the difference between the confidence levels offered by cloud and fog servers can be very high (e.g., as high as 58.6\%).   
A weighted Kantorovich-Rubinstein (K-R) metric is then introduced to quantify the performance difference between the confidence levels of various MNO networks, taking into consideration the heterogeneity in the demands and priorities of different services. 
We then formulate the MNO selection and server adaptation problem as a Markov decision process. To capture short-term and long-term factors in our optimization, 
we investigate two scenarios: (1) {\em finite-horizon decision making}, in which the main objective of a vehicular user is to maximize the average confidence over a given forecasting window, and 
(2) {\em infinite-horizon decision making}, where the user aims at maximizing 
the long-term confidence. We propose a simple threshold policy for deciding 
when and where to switch between MNO networks based on the weighted K-R distance. 
We prove that the proposed threshold policy achieves the optimal performance with low computational complexity. Extensive simulations are conducted to evaluate the performance of AdaptiveFog. 
Numerical results show that AdaptiveFog achieves 30\% to 50\% improvement in the confidence level for fog/cloud latencies, especially when applied to applications with stringent latency requirements (e.g., active road safety applications). 

\section{Related Work}
The concept of fog computing and its relation to 
cloud and mobile edge computing can sometimes be blurry. 
In this paper, we use the term {\em fog computing} to refer to a generalized architecture that includes cloud, edge, and clients\cite{Chiang2017FogBook}. We also use terms {\em fog node} and {\em fog server} interchangeably to denote the servers placed at the edge of the network. We use the term {\em cloud server} to denote the high-performance server installed at the CDC.

\noindent
{\bf Fog Computing-assisted Networking Systems:}
Compared to a CDC, a fog node is a cost-effective yet resource-limited computational device\cite{XY2018EHFogComp,Kortoci2019Infocom}. Most previous works focused on developing new methods and architectures to improve the utilization of fog resources. For example, in \cite{Zeng2016FogComputing}, Zeng {\it et al.} studied the task scheduling and resource management problem to minimize the task completion time.
Tong {\it et al.}\cite{Tong2016EdgeCloud} proposed a hierarchical architecture to improve the resource utilization of a fog computing system. Yu {\it et al.}\cite{Yu2018FogComputing} considered the application provision problem under bandwidth and delay requirements in a fog-enabled Internet-of-Things (IoT) system. Garcia-Saavedra {\it et al.}\cite{GarciaSaavedra2018MEC} proposed an analytical framework, called FluidRAN, that minimizes the aggregated operator expenditure by optimizing the design of the virtualized radio access network. Inaltekin {\it et al.}\cite{Inaltekin2018Fog} introduced an analytical framework to derive the optimal location of the virtual controller for balancing latency and reliability in a fog computing system. In \cite{Deng2016FogCompu}, Deng {\it et al.} derived the optimal workload allocation solution for fog nodes and CDCs so as  to maximize the utilization of the computational resources. \blu{Joint resource allocation for different network entities, such as cloud, fog, and IoT users, has also been investigated using hierarchical frameworks\cite{yuenChau2020FogIoT,XY2017IoTThreeTie}.}

\noindent
{\bf Fog Computing-supported Smart Vehicles:}
Connected vehicle has recently been promoted by both industry and standardization bodies as a key enabler of 
emerging smart vehicule applications, such as intelligent driver assistance and autonomous driving\cite{Zhang2017MECVehicular, Sabella2017MECVehicular}. 
Premsankar {\it et al.}\cite{Premsankar2018MECVehicular} studied the placement of edge computing servers for vehicular applications. An effective heuristic method was proposed to deploy fog servers based on knowledge of road traffic within each deployment area. Lee {\it et al.}\cite{Lee2018V2IMobicom} proposed an in-kernel TCP scheduler to mitigate the network latency of connected vehicles with redundant transmissions. \blu{Joint optimization of communication and computational resources has been studied in connected autonomous vehicles as well as UAVs \cite{yuenChau2020CommCompOptiAutoV,yuenChau2020AcciAutoV,Yuenchau2020V2X,yuenchau2020UAV}.}  

\noindent
{\bf Performance Evaluation:} 
There have been quite a few studies on the performance of vehicular networks supported by a wireless infrastructure. For instance, Bedogni {\it et al.}\cite{Bedogni2018V2V} analyzed a real-world GPS trajectory dataset to investigate the temporal topology of vehicle-to-vehicle (V2V) networks. In \cite{Asadi2018V2I}, Asadi {\it et al.} studied  beam selection for 5G mmWave-based vehicular-to-infrastructure (V2I) communications. An online learning algorithm with environment-awareness was developed and shown to be near-optimal.

In \cite{Mir2014LTEWiFiforVeh}, Hameed Mir {\it et al.} compared the performance of IEEE 802.11p and LTE for vehicular systems using NS-3 simulators. Their results show that LTE offers much better network capacity and performance than IEEE 802.11p. Xu {\it et al.}\cite{Xu2017LTEVehicular} conducted extensive real-world testing of several smart vehicular application scenarios. 
Their results suggest that existing cellular systems are not ideal for  active road safety applications with high-data rate and real-time requirements (e.g., collision avoidance), but sufficient for non-safety-related applications (e.g., traffic updates, file download, Internet access). In \cite{Hadzic2017MECePC}, Hadzic {\it et al.} 
investigated the latency between a fixed mobile station and an LTE-based fog computing system. The authors conducted in-lab testing using an isolated base station with controlled parameters. The results reveal that the wireless connection between the UE and the base station introduces 
non-negligible latency. 

\noindent
{\bf Our Contributions:}
To the best of our knowledge, this is the first work that uses long-term city-wide measurements to model and optimize the latency of a fog computing-based vehicular system. We introduce a novel distance metric, referred to as weighed K-R distance, to quantify the difference in the latency probability distributions of different LTE networks. 
The proposed metric is shown to be a more useful metric than the mean and standard deviation (STD). Accordingly, we derive simple threshold policies for switching between LTE providers and fog/cloud servers while driving. 
Our solution 
can be applied to other technologies, 
including 5G and B5G-based vehicular systems with diverse choices of wireless access technologies and computational resources.  



\section{System Model}
We consider a 
connected vehicular system 
that consists of the following elements: 
\begin{itemize}
\item[-] {\bf UE}: corresponds to a moving vehicle with installed  ITS applications that generate computationally intensive workload requests. Such requests cannot all be handled by  
    the vehicle's onboard computers/processors. The UE may also represent a smart device, 
    a sensor, or a 
    laptop located inside the vehicle.

\item[-] {\bf Wireless Access Networks}: provide wireless links that connect UEs to fog nodes and to the cloud server. In this paper, we consider multi-operator LTE connections in which the UE 
can switch between different MNOs on a per-request basis. 
This is the case, for example, in dual-SIM smart phones that can use two SIM cards to switch between two MNOs, e.g., 
Google's Project Fi-enabled smart phones already have the capability to dynamically switch between different MNO networks\cite{GoogleProjectFi,XY2019MultiOpt}.  

\item[-] {\bf Fog Nodes}: correspond to servers deployed at the edge of the network to support low-latency applications for connected smart vehicles.



\item[-] {\bf Cloud Server}: corresponds to a 
    high-performance server located at the CDC 
to provide on-demand computational services for UEs.
\end{itemize}

Fog nodes can either  be deployed by MNOs as part of their 
network infrastructure or by third-party service providers. 
%
In this paper, we mainly focus on a vehicle-to-cloud/fog system supported by LTE-based 
MNO networks. 

\section{Methodology}
\begin{figure}
\center
\includegraphics[width=3.5 in]{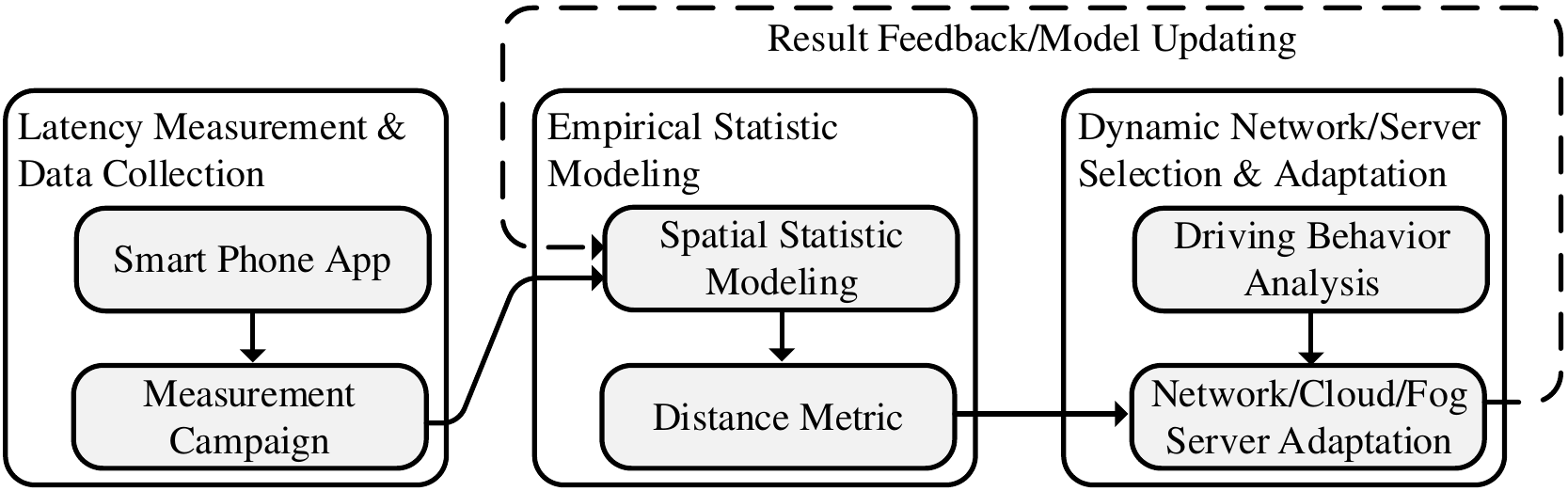}
\caption{\blu{Main components of AdaptiveFog.}} 
\label{Figure_AdaptiveFog}
\end{figure}
\begin{figure}
\begin{minipage}[t]{0.8\linewidth}
\center
\includegraphics[width=2.2 in]{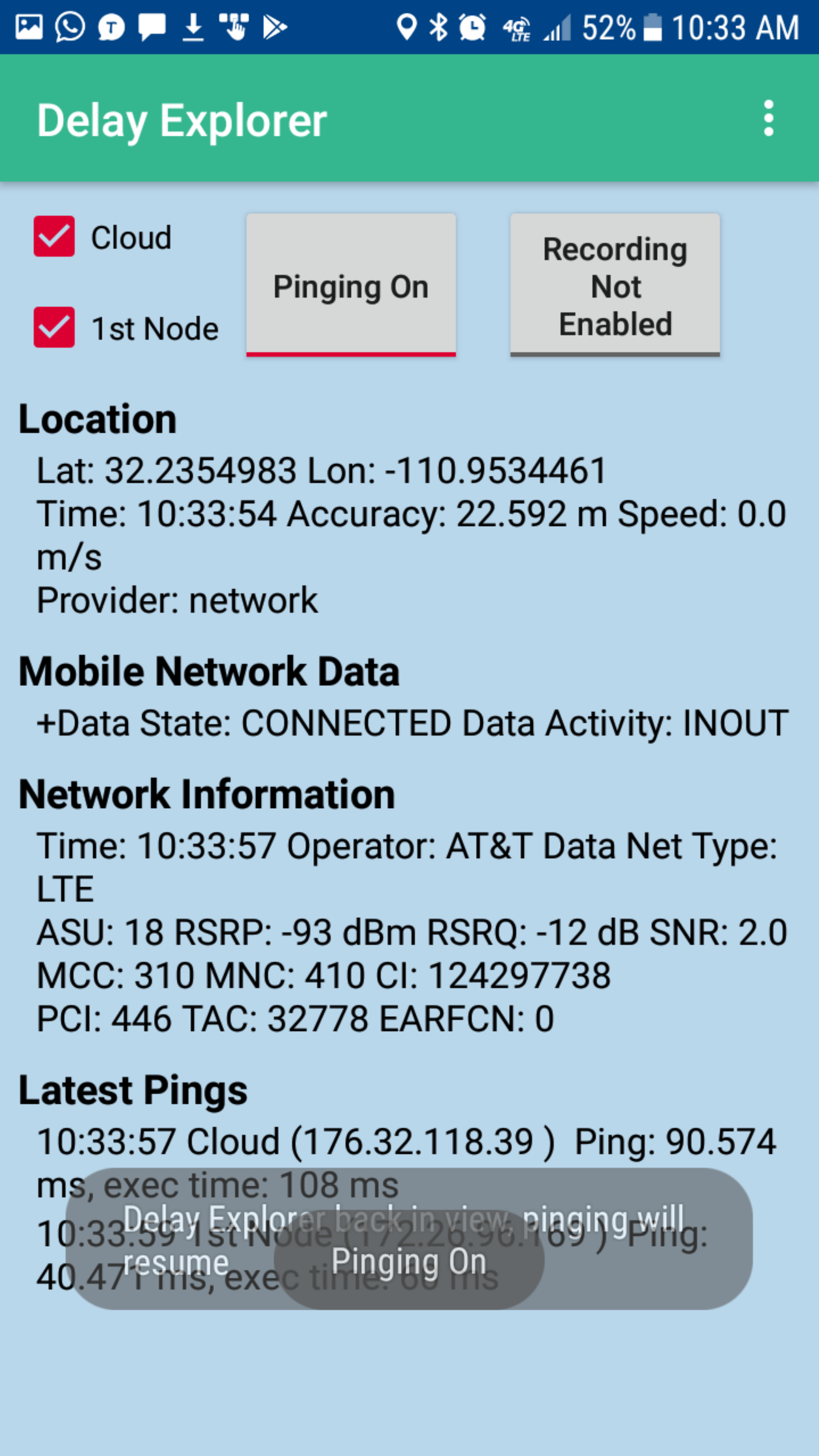}
\caption*{(a)}
\end{minipage}
\begin{minipage}[t]{0.8\linewidth}
\center
\includegraphics[width=3.4 in]{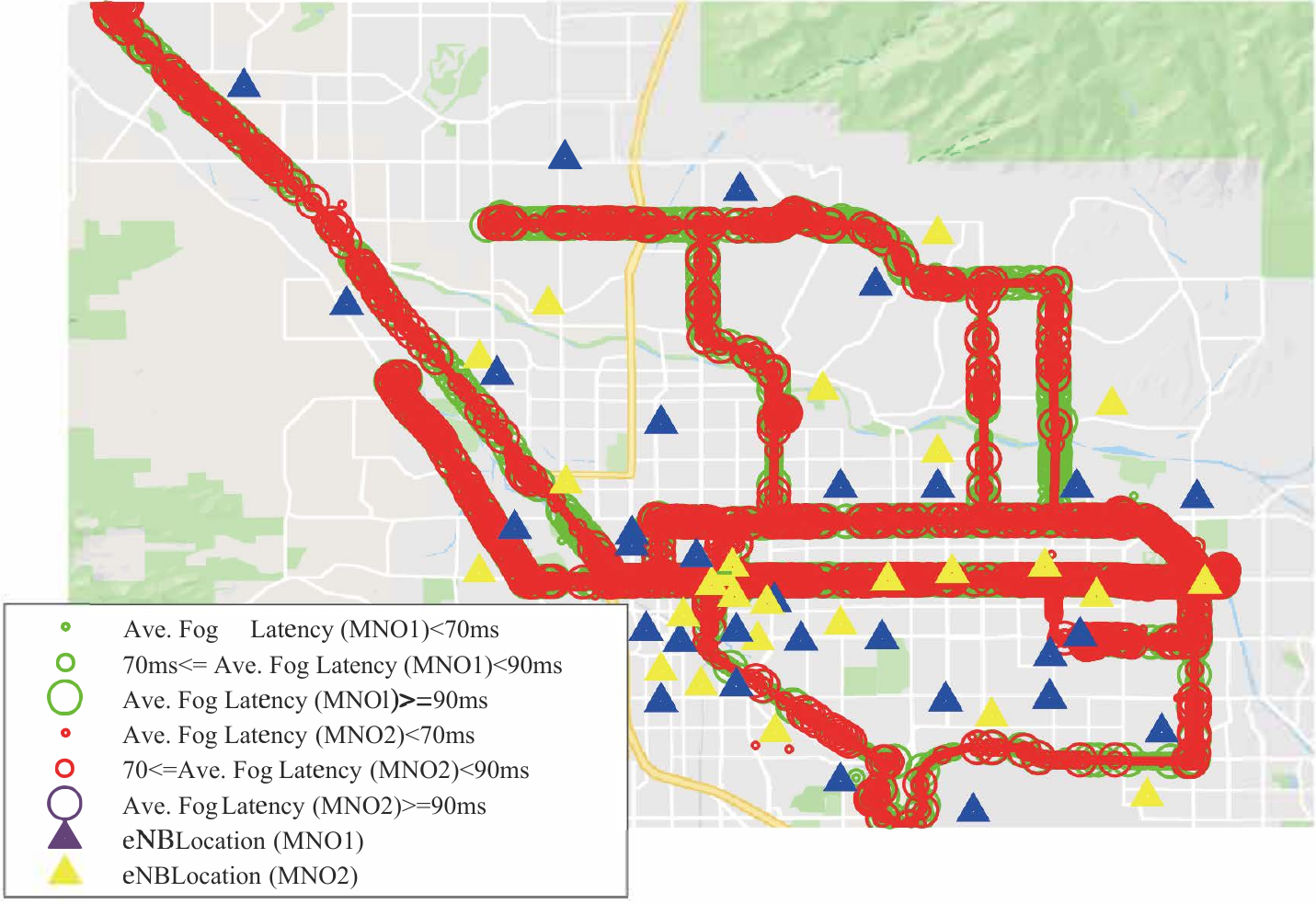}
\caption*{(b)}
\end{minipage}
\caption{(a) User interface of {\em Delay Explorer} developed for our latency measurement campaign, and (b) measurement routes and traces in our study.} 
\label{Figure_Route}
\vspace{-0.1in}
\end{figure}

We propose {\em AdaptiveFog}, a simple framework for the UE to dynamically switch between available MNO networks and cloud/fog servers.
%
AdaptiveFog consists of three main components: trace collection, empirical modeling, and network adaptation, as illustrated in Figure \ref{Figure_AdaptiveFog}.


\subsection{Latency Measurements and Data Collection}
\noindent
{\bf Smart Phone App Design:}
We begin by collecting traces to measure the network performance between a commercial off-the-shelf smart phone located in a moving vehicle and the most likely fog node location as well as the closest cloud server of a major CDC provider. According to \cite{5GAACV2XUserCase}, most safety-related connected vehicular services, such as lane-change assist and danger/hazard 
warning, require timely and reliable delivery of small information packets mostly with sizes between 500 and 1000 bytes. In these applications, latency is often considered more important 
than other traditional metrics such as channel capacity and throughput. Therefore, instead of measuring the network bandwidth, we evaluate the RTT. \blu{We use a dedicated smart phone app, called {\em Delay Explorer}, to periodically ping the IP address of the most likely fog node location and an arbitrary IP address of the closest cloud server of a major CDC. Delay Explorer was developed using Android API. We fix the packet size to 500 bytes and record the resulting RTT every 500 ms.} 
In addition to recording the RTT, Delay Explorer also records other vehicle-related information such as timestamp, GPS coordinates, altitude, driving speed, as well as network-related information, including network connection type, bearing, ASU, RSRP, RSRQ, etc., as shown 
in Figure \ref{Figure_Route}(a). 
Delay Explorer only records data when the phone is connected to an LTE network; it stops recording if the LTE connection is dropped.

\noindent
{\bf Measurement Campaign:} We conducted a five-month city-wide measurement campaign with a Google Pixel 2 smart phone (e.g., UE) running the Delay Explorer app. For the first month, the phone was placed at different but  fixed locations across a university campus and in a residential area for continuous recording. For the remaining 4 months, it was placed inside a vehicle for driving measurements during the rest 4 months (see Figure \ref{Figure_Route}(b) for the measuring routes). We purchased SIM cards from two major MNOs, Sprint and AT\&T, and collected equivalent amounts of latency samples from both MNOs during both fixed and driving measurements. 
For each MNO, the UE records two types of latency: 1) {\it cloud latency}, which represents the RTT recorded from pinging the IP address of a cloud server, and 2) {\it fog latency}, which corresponds to the RTT recorded from pinging the first hop IP address in each MNO's LTE network.
For the driving measurements, the vehicle records latency traces for 
approximately 2 hours per day. We have collected over 300,000 RTT values from each MNO's network while driving along the main routes in the city of Tucson, Arizona.

\begin{figure}%
\begin{minipage}[t]{0.5\linewidth}
\centering
\includegraphics[width=1.8 in]{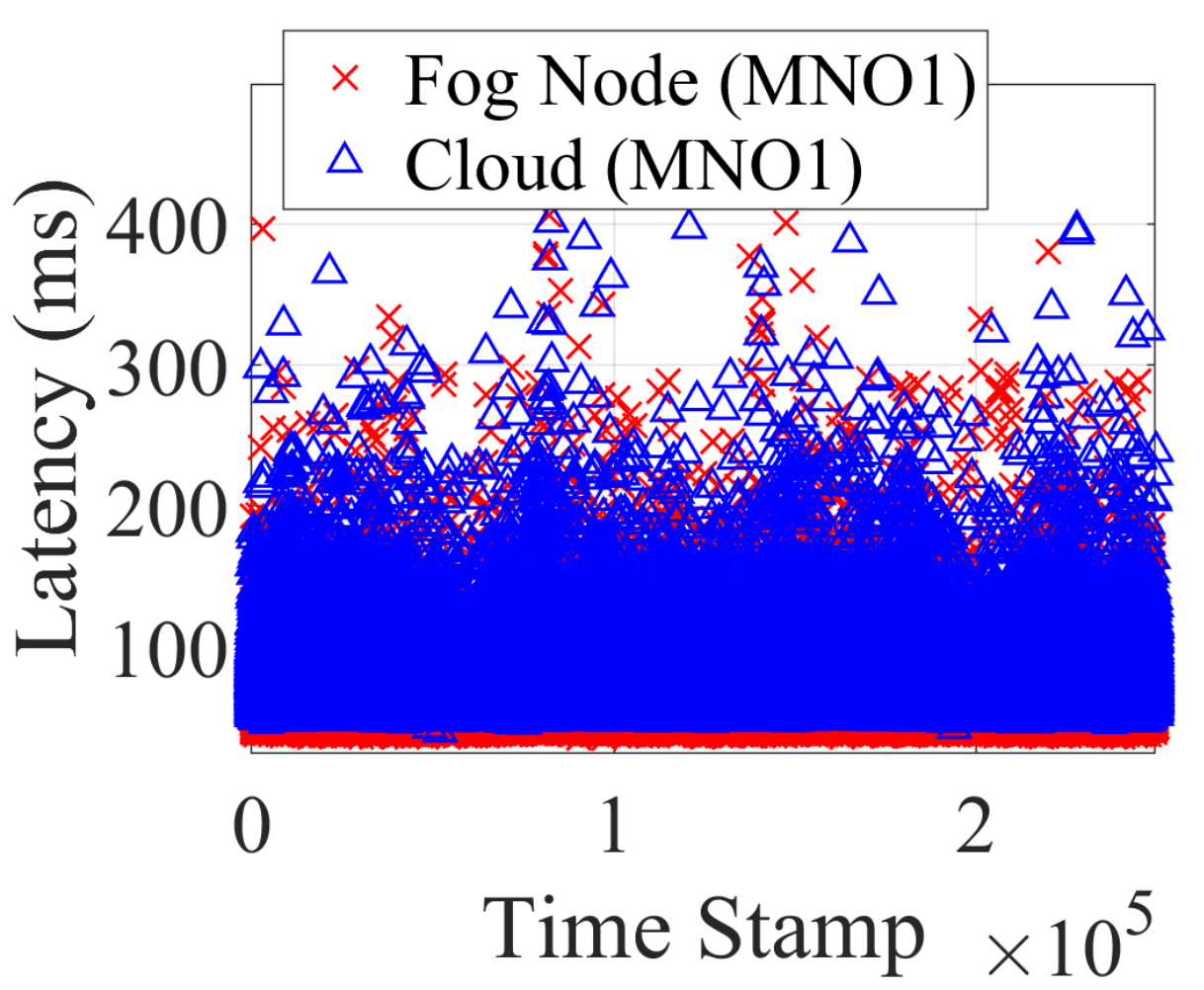}
\vspace{-0.1in}
\captionsetup{labelformat=empty}
\caption*{(a)} 
\label{Figure_PDFCloudLatencyFixLoc}
\end{minipage}
\begin{minipage}[t]{0.45\linewidth}
\centering
\includegraphics[width=1.9 in]{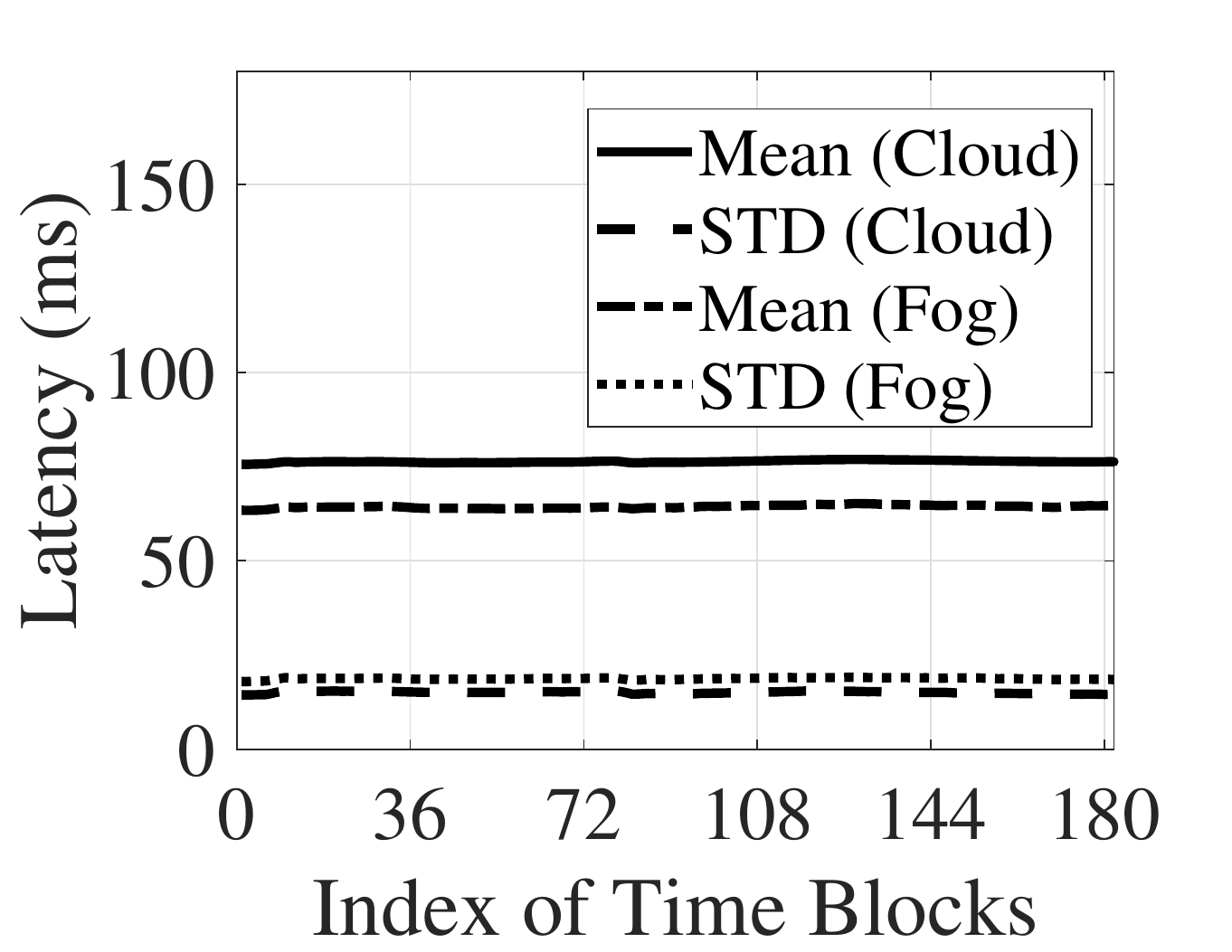}
\vspace{-0.1in}
\captionsetup{labelformat=empty}
\caption*{(b)} 
\label{Figure_PDFCloudLatencyFixLoc}
\end{minipage}
\vspace{-0.1in}
\caption{(a) Latency trace at a fixed location in a university campus, and (b) corresponding mean and STD recorded with 1 hour windows for a one-week continuous measurement.}
\label{Figure_FixLatencyVSTime}
\vspace{-0.1in}
\end{figure}

\begin{figure}
\begin{minipage}[t]{0.5\linewidth}
\centering
\includegraphics[width=1.8 in]{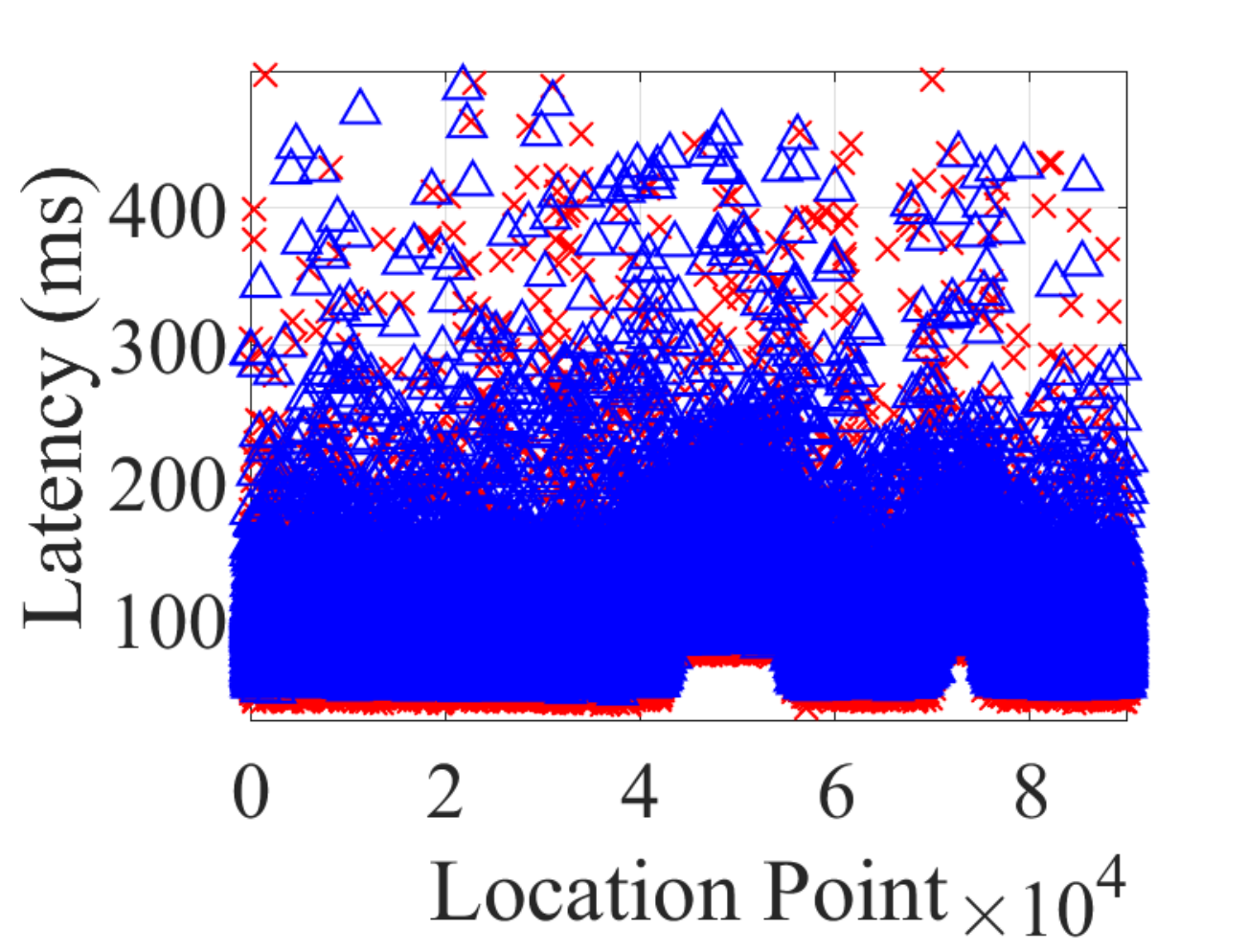}
\captionsetup{labelformat=empty}
\caption*{(a)}
\label{Figure_PDF1stNodeLatencyFixLoc}
\end{minipage}
\begin{minipage}[t]{0.45\linewidth}
\centering
\includegraphics[width=1.9 in]{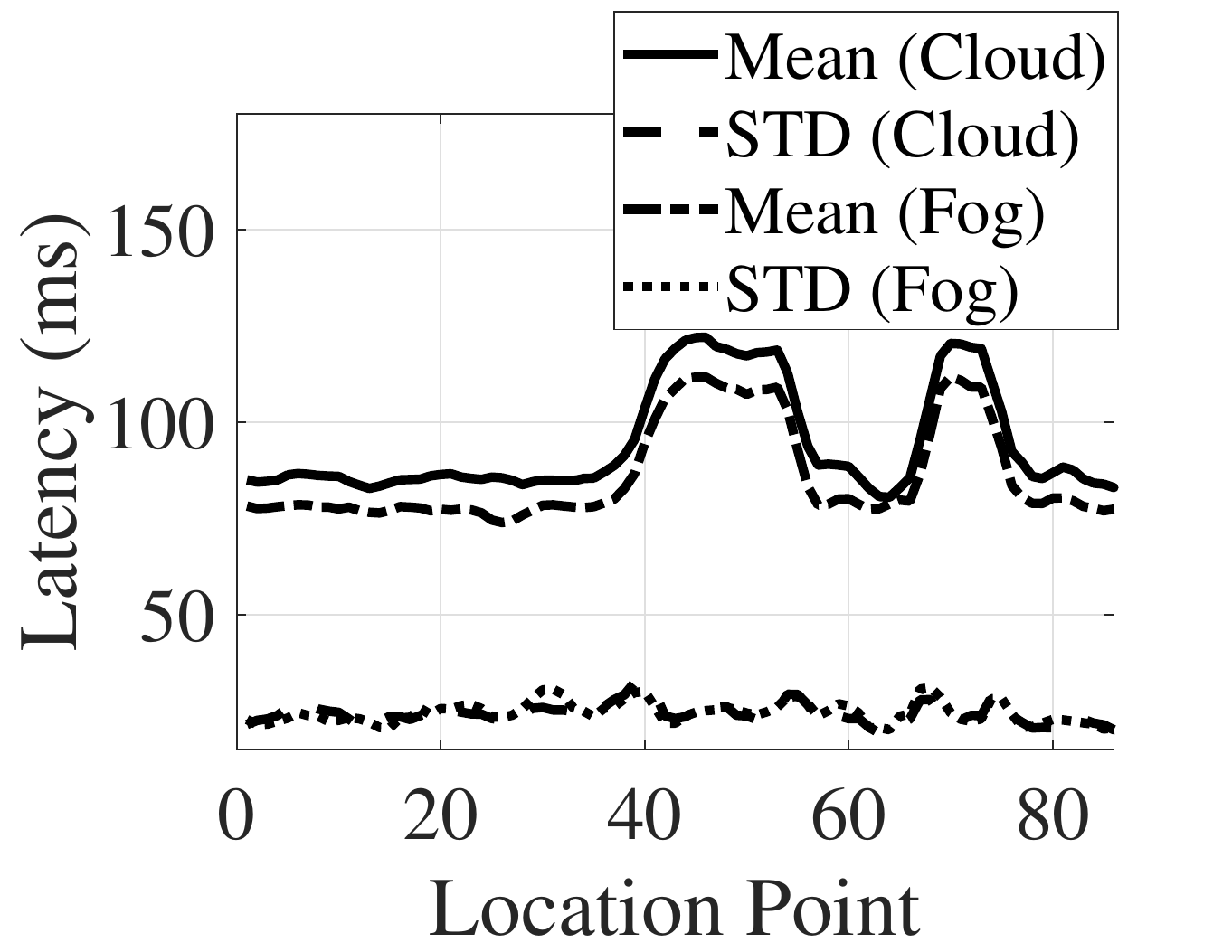}
\captionsetup{labelformat=empty}
\caption*{(b)}
\label{Figure_PDF1stNodeLatencyFixLocAveSTD}
\end{minipage}
\vspace{-0.1in}
\caption{(a) Trace recorded on a main road in the city center of Tucson, AZ, and (b) the corresponding mean and STD.}
\label{Figure_DrivingLatencyVSDiffLoc}
\end{figure}

\begin{table*}[]
\center
\caption{\blu{Statistics of collected RTT samples over two MNO networks in Tucson, Arizona}}
\begin{tabular}{|l|l|l|c|c|c||c|c|c|}
\hline
\multicolumn{3}{|c|}{Traces}                               & \begin{tabular}[c]{@{}c@{}}L1\\ Fixed\end{tabular} & \begin{tabular}[c]{@{}c@{}}L2\\ Fixed\end{tabular}  & \begin{tabular}[c]{@{}c@{}}All\\ Fixed\end{tabular} & \begin{tabular}[c]{@{}c@{}}R1 (Driving) \\ (6.1m/s)\end{tabular} & \begin{tabular}[c]{@{}c@{}}R2 (Driving)\\ (15.7m/s)\end{tabular} & \begin{tabular}[c]{@{}c@{}}All\\ Driving\end{tabular} \\ \hline
\multirow{8}{*}{MNO 1} & \multirow{4}{*}{\begin{tabular}[l]{@{}l@{}}Fog\\ Latency\\(ms)\end{tabular}}   &Mean      & 62 & 72  & 70                                                  & 83                                                            & 96                                                            & 88                                                  \\ \cline{3-9}
                      &                        & STD       & 18 & 16  & 18                                                  & 28                                                            & 29                                                            & 34                                                  \\ \cline{3-9}
                      &                        & Median    & 55 & 71  & 68                                                  & 77                                                            & 91                                                            & 85                                                  \\ \cline{3-9}
                      &                        & Conf. 90\% & 85 & 86  & 85                                                  & 115                                                           & 121                                                           & 120                                                 \\ \cline{2-9}
                      & \multirow{4}{*}{\begin{tabular}[l]{@{}l@{}}Cloud\\ Latency\\(ms)\end{tabular}} & Mean      & 74 & 87  & 85                                                  & 94                                                            & 108                                                           & 96                                                  \\ \cline{3-9}
                      &                        & STD       & 15 & 15  & 21                                                  & 26                                                            & 29                                                            & 33                                                  \\ \cline{3-9}
                      &                        & Median    & 71 & 88  & 86                                                  & 92                                                            & 108                                                           & 94                                                  \\ \cline{3-9}
                      &                        & Conf. 90\% & 88 & 100 & 104                                                 & 124                                                           & 129                                                           & 128                                                 \\ \hline \hline
\multirow{8}{*}{MNO 2} & \multirow{4}{*}{\begin{tabular}[l]{@{}l@{}}Fog\\ Latency\\(ms)\end{tabular}}   & Mean      & 72 & 64  & 72                                                  & 85                                                            & 80                                                            & 83                                                  \\ \cline{3-9}
                      &                        & STD       & 14 & 17  & 15                                                  & 52                                                            & 46                                                            & 51                                                  \\ \cline{3-9}
                      &                        & Median    & 71 & 93  & 71                                                  & 69                                                            & 67                                                            & 66                                                  \\ \cline{3-9}
                      &                        & Conf. 90\% & 84 & 87  & 86                                                  & 132                                                           & 112                                                           & 131                                                 \\ \cline{2-9}
                      & \multirow{4}{*}{\begin{tabular}[l]{@{}l@{}}Cloud\\ Latency\\(ms)\end{tabular}} & Mean      & 87 & 74  & 88                                                  & 119                                                           & 125                                                           & 124                                                 \\ \cline{3-9}
                      &                        & STD       & 13 & 13  & 17                                                  & 50                                                            & 47                                                            & 54                                                  \\ \cline{3-9}
                      &                        & Median    & 88 & 71  & 90                                                  & 108                                                           & 117                                                           & 109                                                 \\ \cline{3-9}
                      &                        & Conf. 90\% & 99 & 87  & 102                                                 & 166                                                           & 133                                                           & 100                                                 \\ \hline
\end{tabular}
\label{Table_LatencyPerformance}
\vspace{-0.2in}
\end{table*}



\noindent
{\bf Results and Discussion:}
In Figure \ref{Figure_FixLatencyVSTime}(a), we present the trace of one week of recorded latencies at a fixed location (a university lab). 
We observe that RTT samples often fluctuate between consecutive measurements and there are no observable correlation between the instantaneous RTT and the time-of-measurement. Latency statistics, including the mean and standard deviation (STD), at a given location are surprisingly stable (see Figure \ref{Figure_FixLatencyVSTime}(b)). This result is consistent with a recent study in a similar-sized city\cite{Lee2018V2IMobicom}.
To investigate the impact of different locations on service latency, we aggregate all the driving traces collected along a major route across the city center during the four-month period and present the RTT samples recorded at different location points in Figure \ref{Figure_DrivingLatencyVSDiffLoc}. We observe noticeably different patterns in some locations than others. In other words, compared to the time of measurements, the geographical heterogeneity contributes more to the diversity in the RTT statistics. 
%
%
\blu{In Table \ref{Table_LatencyPerformance}, we present the key statistics of collected RTT samples, including the mean, STD, and median, at two fixed locations (labeled as ``L1 fixed" and ``L2 fixed") and for two driving routes (labeled as ``R1 (Driving)" and ``R2 (Driving)" at average driving speeds of 6.1 m/s and 15.7 m/s, respectively). We also present the combined RTT statistics for the fix and driving scenarios (labeled as ``All Fixed" and ``All Driving", respectively). It can be observed that the RTTs experienced over the two different MNO networks can vary significantly in some locations/driving routes. Neither MNO shows consistent advantage over the other across all the service area in terms of latency performance.} When taking into consideration of all the traces, both MNOs exhibit similar latency performance in terms of mean and STD at a fixed location. However, the driving traces of different MNOs exhibit more noticeable differences in terms of STD, mean, and median values. One of the main reasons for this behavior is that the eNB deployment densities and locations of different MNOs can be quite different, as shown in Figure \ref{Figure_Route}(b). We will give a more detailed discussion about various issues that can affect the latency 
in Section \ref{Section_LTELatency}.

\subsection{Model Evaluation}

\noindent
{\bf Weighted Confidence:}
Most latency-sensitive vehicular applications do not observe noticeable performance difference as long as the resulting RTT is below their tolerable thresholds\cite{5GAACV2XUserCase}. For example, it is reported in \cite{Mir2014LTEWiFiforVeh} that for {\em active road safety applications}, such as collision avoidance, emergency alert, and active control intervention for crash prevention, the maximum tolerable service latency is around 100 ms. For {\em cooperative traffic efficiency applications} 
which are used to improve the traffic flow and enhance the traffic coordination, 
less than 200 ms of latency is considered as sufficient. For most {\em infotainment applications}, up to 500 ms of latency is tolerable.

We consider the {\em proportionally weighed confidence level} as the main metric for evaluating the latency performance of a connected vehicular system. This metric is formally introduced as follows. Suppose that each UE can request a finite set of services, denoted as $\cal M$, each with its own maximum tolerable latency, denoted as $r_i$ for service type $i$. 
The {\em confidence level} $F_i$ of service type $i$ is the probability that maximum tolerable latency $r_i$ can be satisfied: 
\begin{eqnarray}
F_i = \Pr \left( x \le r_i \right) 
\end{eqnarray}
where $x$ is the latency.
%
\blu{The confidence level can be directly calculated from a mathematical model, e.g., a probability distribution function that estimates real latency values. However, recent results in \cite{Hadzic2017MECePC, Lee2018V2IMobicom} as well as our own measurements indicate that the observed latency in modern cellular systems is too complex to be accurately captured by any well-known probability distribution function nor it can be directly derived via a simple mathematical equation. Compared to existing works, which first formulate mathematical models and then present the corresponding simulation results, in this paper, we take a different approach where we first conduct extensive measurements over commercial cellular networks deployed by two major MNOs and then analyze each of the possible factors that impact latency, including fog node placement, uplink/downlink transmission, handover, and driving behavior. Finally, we establish an empirical probability distribution of the confidence level $F_i$ for each supported service $i$. Note that for each specific service with a given $r_i$, $F_i$ will have a fixed value at each location.}

Compared to traditional metrics, such as the mean, minimum, and instantaneous latency, the confidence level is a more realistic and practical metric,  because, for most vehicular applications,  
it is critical to quantify the chance that a certain latency threshold can be guaranteed by the wireless access network. Furthermore, as we show later, although the difference between the mean values of fog and cloud latency can be as low as 10 to 20 ms (see Table \ref{Table_LatencyPerformance}), the difference between their confidence levels can be significant, e.g., as high as 58.6\% (see Section \ref{Section_MeasureAnalysis}).

Different types of services can have different probabilities of being requested as well as priorities to be served. For example, cooperative traffic efficiency applications may be requested more often in low-speed traffic congestion areas compared to active road safety applications, which are 
assigned with a higher priority compared to infotainment applications. To include these factors into the latency performance analysis, we assign each service type $i$ with a weighting factor $w_i$. The proportionally weighed confidence level $\hat F$ is then defined as the aggregated confidence levels with all the supported services being served at their corresponding tolerable latency thresholds:
\begin{eqnarray}
\hat F = \sum_{i\in {\cal M}} w_i F_i.
\label{eq_WeightedConfidenceLevel}
\end{eqnarray}


Note that (\ref{eq_WeightedConfidenceLevel}) 
is a general performance metric that can be applied to a wide range of  applications under various scenarios. For example, by settling $w_i$ to the probability that type $i$ service will be requested, 
$w_i F_i$ is equivalent to the probability that service $i$ is requested by the UE is served with satisfactory latency performance. 


\noindent
{\bf Distance Metric:}
To quantify the difference between the latency performance for different wireless access networks (e.g., MNO networks), we introduce the weighted Kantorovich-Rubinstein (K-R) distance,   
defined as: 
\begin{eqnarray}
K(F, G) = \sum_{i\in {\cal M}} w_i \left( F_i - G_i \right)
\label{eq_KRDistance}
\end{eqnarray}
where $F_i$ and $G_i$ are the empirical cumulative distribution functions (CDFs) of latency offered by  two different MNOs. 

The weighted K-R distance in (\ref{eq_KRDistance}) corresponds to the weighted difference between the confidence levels of different services at their maximum tolerable thresholds. Generally speaking, the UE should always choose the LTE network that provides a higher confidence level to achieve a better latency performance guarantee. However, there is a cost for switching between LTE networks. This cost is related to the extra price paid to multiple MNO networks, extra latency for the UE to disconnect from one MNO and reconnect to another, and/or extra energy and processing resource consumed during the switching. Therefore, the UE needs to not only consider the current performance of each MNO but also the performance that can be offered by the MNOs in the future as well as the switching cost, i.e., the UE should choose a single MNO or a sequence of MNOs to maximize the confidence of maintaining the guaranteed services at the minimal cost incurred when switching back-and-forth between different  networks.

\noindent
{\bf Model Updating:} The probability distribution of the latency in some specific locations can be affected by some unexpected events, e.g., road work and/or traffic accidents. In this case, the UE should also be able to detect the change and adjust the empirical PDF according to the updated latency traces. Several approaches can be applied to detect the change of empirical PDF using updated samples\cite{Kifer2004PDFChange}. Applying and comparing the model/statistic-changing detection methods in AdaptiveFog is out of the scope of this paper and will be left for future research. 
\subsection{Network/Server Selection and Adaptation}
\noindent
{\bf Driving Behavior Modeling:}
In addition to the network infrastructure, the latency performance of the UE 
can also be affected by human-related factors such as the driving routine,  and driver's behavior. Many existing works as well as our own measurements indicate that the driving location and speed of a vehicle typically follow a Markov behavior, that is the future state of the vehicle including its location and speed only depends on the current state. 
We apply the driving location and speed data collected in our measurement campaign to calculate the empirical state transition probability of the UE when driving through different locations with different speeds. 

\begin{figure}
\centering
\includegraphics[width=2.8 in]{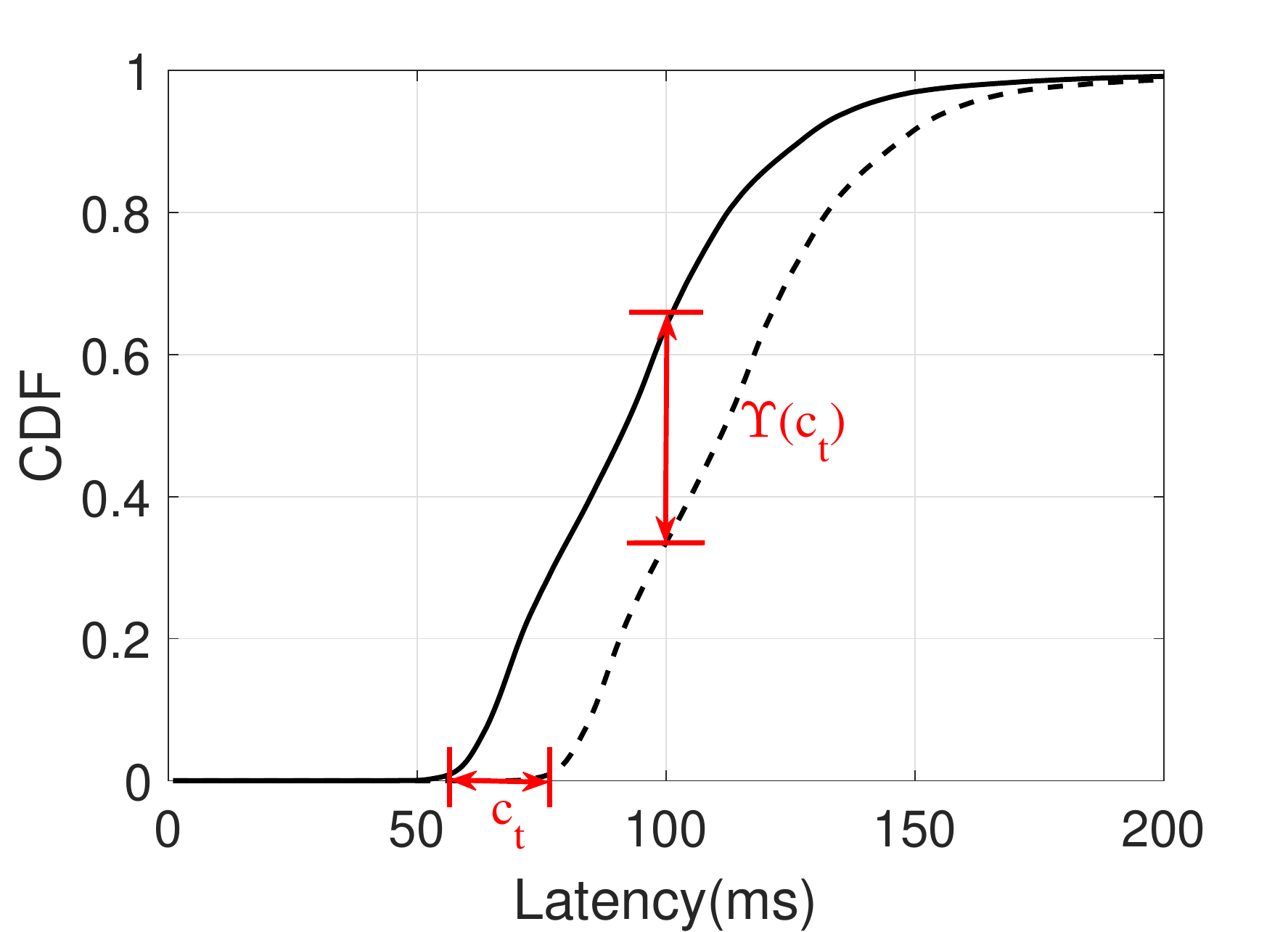}
\caption{\blu{\small Switching cost $\Upsilon \left(c_t\right)$ where  $c_t$ is the extra latency caused by switching from one network (with CDF in dashed
line) to another (with CDF in solid line).}} 
\label{Figure_SwitchingCost}
\end{figure}

\noindent
{\bf Network Adaptation:}
The main objective of our network adaptation design is to maximize the long-term confidence minus the possible cost incurred due to MNO switching while the UE is driving through different locations. We consider a slotted decision making process and assume in each time slot $t$, the UE can only choose one MNO's network. With some abuse of the notation, we use $k$ to denote the selected MNO as well as its LTE network. We also use $j$ to denote the fog or cloud server selected by the UE. As will be shown in Section \ref{Section_MeasureAnalysis}, the cloud latency is often larger than the fog latency. However, a cloud server has much more computational resources compared to a fog server and therefore can still be considered as the preferred choice for workload outsourcing if the latency requirement is not stringent. \blu{We write the utility obtained by the UE in time slot $t$ as follows:
\begin{eqnarray}
u_t(k_t,j_t) = \sum\limits_{i\in {\cal M}} {w}_i F_{i,t} \left( s_{t}, k_t, j_t \right)- {\bf 1} \left( k_t\neq k_{t-1} \right) \Upsilon \left(c_t\right)
\label{eq_InstantUtility}
\end{eqnarray}
where we use subscript $t$ to denote the parameters in time slot $t$. ${\bf 1} \left(\cdot \right)$ is the indicator function. $\Upsilon \left(c_t\right)$ is the reduction in the confidence level caused by switching between different wireless access networks, where $c_t$ is the cost of switching in terms of the extra latency for the UE to disconnect from one MNO and reconnect to another, the extra energy and processing resource consumed during the switching, and the extra price paid to multiple MNO networks. Figure \ref{Figure_SwitchingCost} illustrates the reduction of confidence level $\Upsilon \left(c_t\right)$ caused by $c_t$.} $s_t$ is the state at time $t$, which includes the location, speed, and currently connected MNO network. $F_{i,t} (s_t, k_t, j_t)$ is the confidence level when MNO $k_t$ is selected by the UE. 

%

To capture the scenarios where a connected vehicle may need to optimize its short-term or long-term latency performance, we investigate two decision making processes at the UE: finite-horizon and infinite horizon. In the first scenario, the UE tries to maximize the confidence level for a given duration of time in the future. For example, the UE may drive through a sequence of complex intersections that requires high confidence of latency guarantee.  In this case, the UE will carefully decide a sequence of MNOs to be connected with when driving into different locations at different times. Suppose the UE tries to maximize its confidence level in the next $T$ time slots of driving. The optimal policy for the UE to select the optimal MNO and fog/cloud server can be written as
\begin{eqnarray}
\pi \left( s_t \right) = \arg \max\limits_{\langle {k}_t, {j}_t \rangle} \mathbb{E} \left( \sum\limits^T_{\tau=t} u_\tau \left(k_\tau,j_\tau\right) \right)
\label{eq_ShortTermUtility}
\end{eqnarray}
where $\mathbb{E} \left(\cdot\right)$ is the expectation. 

In the infinite horizon scenario, the UE focuses on optimizing the confidence level for a long-term driving experience. For example, the UE may not know the final destination or that the route may consist of a very large number of time slots. 
In this case, we can write the optimal policy for the UE to select the optimal MNO and fog/cloud server as
\begin{eqnarray}
\pi' \left( s \right) = \arg \max\limits_{\langle k, j \rangle} \mathbb{E} \left( \lim_{T \rightarrow \infty} \sum\limits^T_{t=1} \gamma^t u_t \left(k,j\right) \right)
\label{eq_LongTermUtility}
\end{eqnarray}
where $0 < \gamma \le 1$ is the discount factor specifying how impatient the UE is, i.e., the smaller the $\gamma$ the more the UE cares about the short-term latency. 
$\gamma$ also ensures the accumulated latency of the UE to be finite.


%
%
%


\section{Latency Analysis in LTE-based Fog Computing}
\label{Section_LTELatency}

The RTT between the UE and the fog node can be affected by the following factors:

\noindent
{\bf Fog Node Placement:}
Most existing works assume that by simply deploying fog servers at or near the eNB (the closest network element to users), one can minimize the RTT between the UE and the fog server\cite{Sabella2017MECVehicular,XY2017FogCompInfocom,XY2018TactileInternet}. However, as observed in \cite{Hadzic2017MECePC}, eNBs are typically installed at inaccessible locations (e.g., the top of a hill, lamp posts, and street cabinets), and therefore cannot offer sufficient space and resources (e.g., electric power and cooling load) for servers. In addition, allowing the computational workload sent by the UE to be redirected to a co-located server at the eNB instead of being forwarded to the higher layer of LTE network, i.e., service-gateway (S-GW) and ePC, via S1 interface will also require a total redesign of LTE interfaces.  
%
%
%
In commercial LTE systems, UE data packets 
pass through many IP routing hops within the ePC\cite{Lee2018V2IMobicom}. 
The addresses of these internal IP hops are hidden from public access. The UE can only get a private subnet IP address that is translated to a public address at the packet-gateway (P-GW). In fact, in our measurements, we observe that in each MNO network, the IP address of the first hop IP address identified by ``traceroute" remains the same across different cities. This is typical for IPv4-based networks, where IP addresses are scarce. 
In this case, an Internet-based application at the UE perceives the entire ePC as a single routing hop.
To minimize the RTT between the UE and the fog node, the fog node should be placed close to the first public IP address, also referred to as the first node in ePC, that can be identified. 

\noindent
{\bf Uplink Latency:}
We consider the scenario where the UE submits its workload using the data-only best-effort service offered by MNOs. In this case, the UE must first initiate the uplink data transmission by submitting a one-bit scheduling request (SR) to the physical uplink control channel (PUCCH), informing the eNB about the upcoming packet arrivals. The UE will then wait for the eNB to schedule a grant that specifies the radio resources for uplink transmission. If the UE does not receive the uplink resources from the eNB, it will resend the SR on PUCCH based on the SR periodicity from 5ms to 80ms (in LTE Release 9, new 1 ms and 2 ms SR periodicities have been added.)\cite{LTEA36331RadioResourceControl}. 

\noindent
{\bf Downlink Latency:} The eNB will feedback the processing result to the UE when it becomes available. In LTE-FDD, a 1 ms subframe 
is considered to be the typical wireless transmission time interval between the UE and the eNB. There is also a frame alignment time (typically 0.5 ms) and UE processing latency (1.5 ms). If the delivery fails, the UE will feed back a negative acknowledgment (NACK) after 4 subframes, which will then trigger a Hybrid ARQ (HARQ) retransmission, bringing the total delay to 8 ms. 





\noindent
{\bf Handover:}
One of the main factors that cause service interruption, drop of connection, and increased latency for the UE is handover between two eNBs. 
The handover decision is typically initiated by the UE via its connected eNB when the measured downlink signal power drops off
below a certain threshold. In particular, the UE starts measuring the signal strength of a neighboring eNB when the received signal power of the current eNB is below the threshold value. The UE will then report the result to the source eNB. Because signal strength measurements and neighboring cell search are conducted by the UE, even in the idle state during DRX periods, their associated latency is negligible. Once the downlink measurement results reported by the UE satisfy certain conditions, the source eNB will initiate the handover process by sending a radio resource control (RCC) reconfiguration message to the UE. This message specifies the identity of the target eNB. According to \cite{3GPPLTE23.009Handover}, the maximum allowed delay for RCC reconfiguration is 15 ms. The source eNB will also send a handover request message to the target eNB. Once it receives the request, the target eNB will allocate the required resources in its cell and will also assign a new Radio Network Temporary Identifier (RNTI) to the UE. 
The handover can be based on the S1 interface between two eNBs without requiring coordination through higher-level components, such as MME and P-GW. If the S1 interface is unavailable, the handover will be processed by the MME via the X1 interface. From the UE's perspective, it is impossible to differentiate these two types of handover. In fact, it is generally impossible for the UE to tell which handover procedure has been executed.

\section{Empirical Modeling}
\label{Section_MeasureAnalysis}

\begin{figure}
\centering
\includegraphics[width=3.5 in]{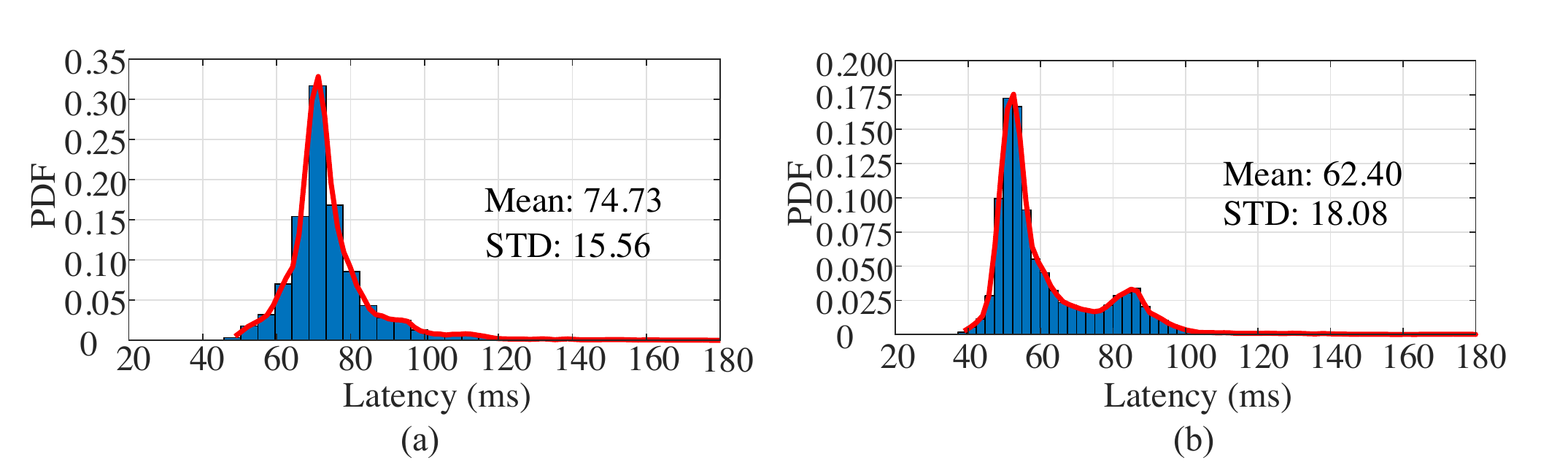}
\vspace{-0.2in}
\caption{\small PDF of (a) cloud and (b) fog latency at a fixed location.} 
\label{Figure_PDFCloudLatencyFixLoc}
\end{figure}


\begin{figure}
\centering
\includegraphics[width=3.5 in]{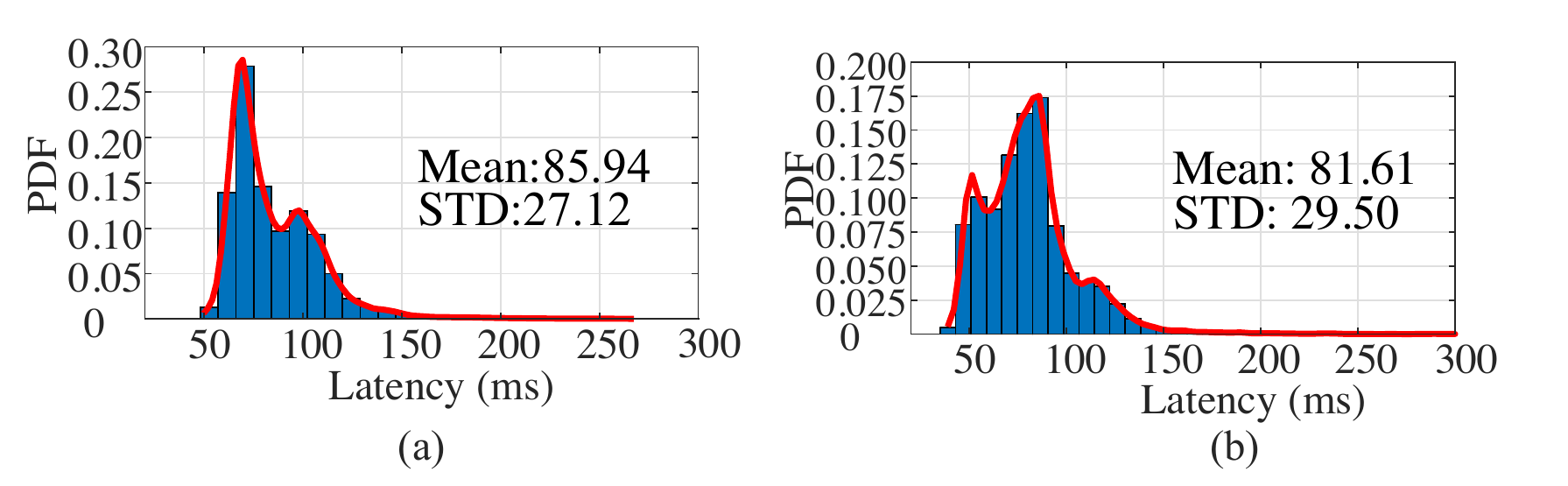}
\vspace{-0.15in}
\caption{\small PDFs of MNO 1's (a) cloud and (b) fog latency during driving.} 
\label{Figure_MNO1PDFLatencyDriving}
\vspace{0.1in}
\includegraphics[width=3.5 in]{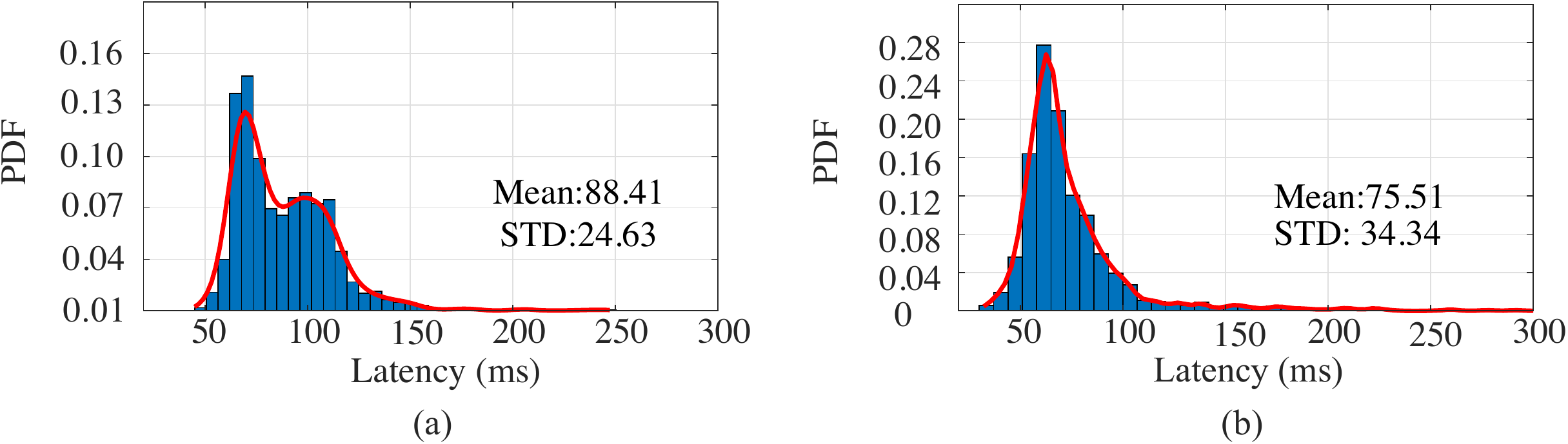}
\vspace{-0.15in}
\caption{\small PDFs of MNO 2's (a) cloud and (b) fog latency during driving.} 
\label{Figure_MNO2PDFLatencyDriving}
\end{figure}
%
%
\begin{figure}[t]
\centering
\includegraphics[width=3.5 in]{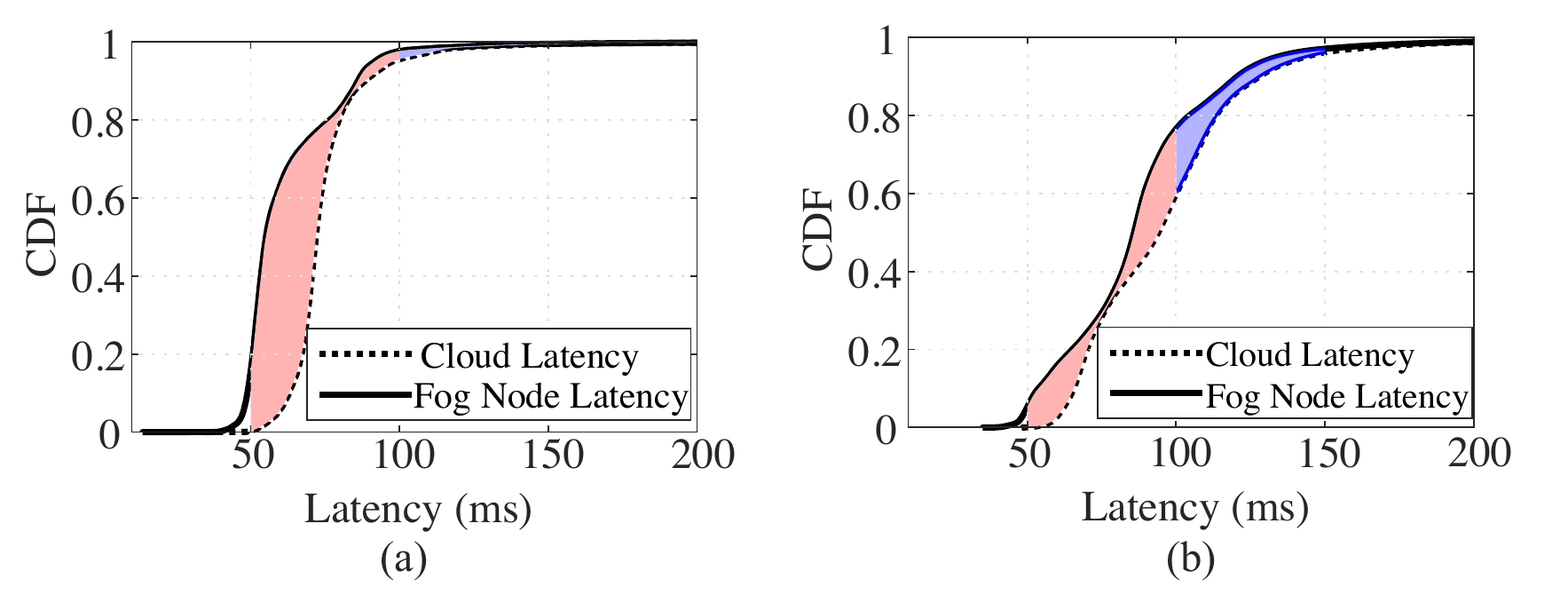}
\vspace{-0.2in}
\caption{\small Latency CDF (a) at a fixed location, and (b) while driving  (maximum tolerable latency thresholds of 50, 100, and 150 ms). The size of the shaded areas in different colors correspond to the K-R distance between cloud and fog latencies with different tolerable thresholds.} 
\label{Figure_KRDistanceFixedandDrivingComp}
\end{figure}



\begin{figure}
\centering
\includegraphics[width=3.5 in]{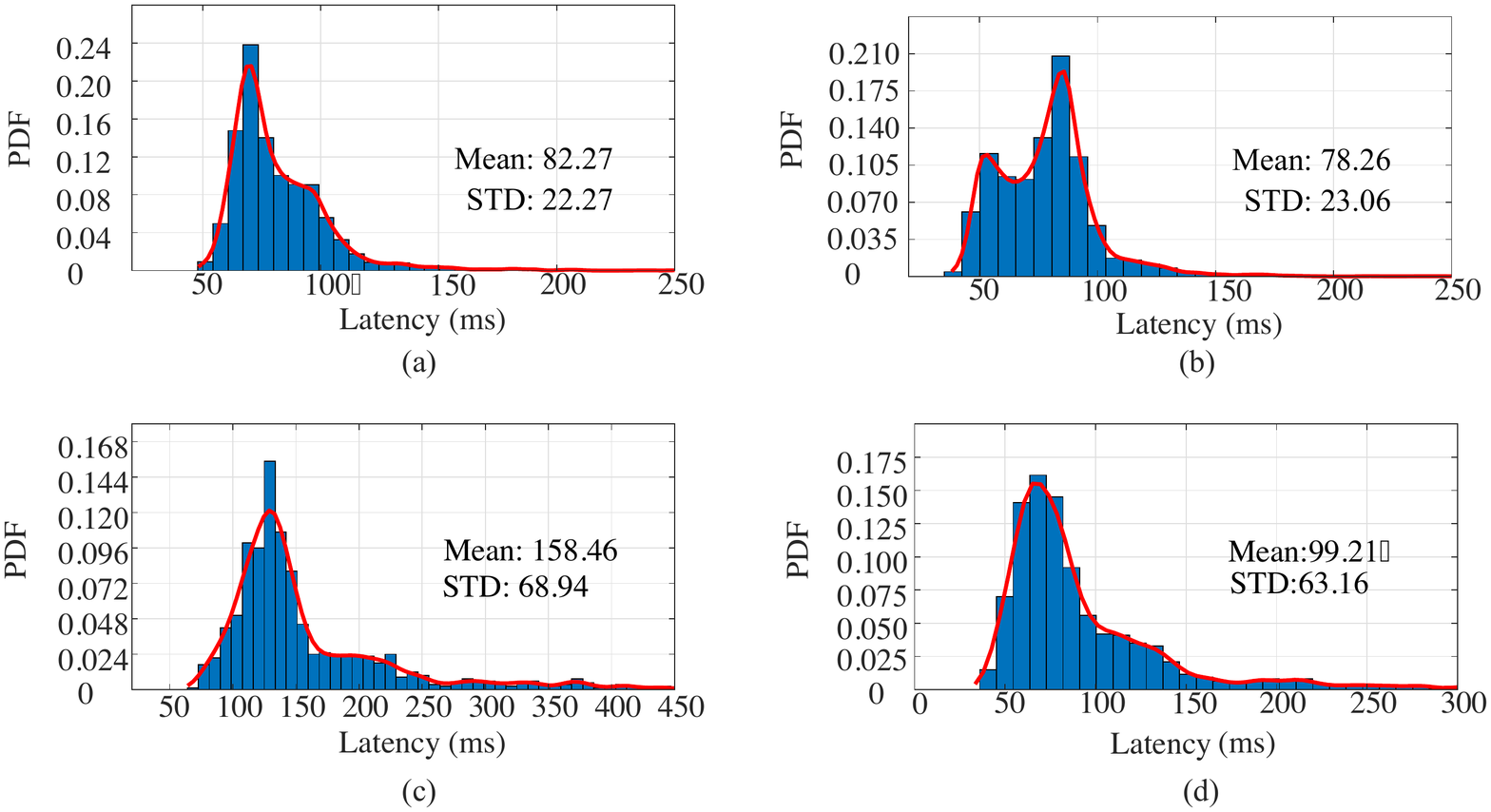}
\vspace{-0.15in}
\caption{\small PDFs of MNO 1's  (a) cloud and (b) fog latencies in a multistory parking lot.} 
\label{Figure_MNO1PDFParkingLot}
\vspace{0.1in}
\includegraphics[width=3.5 in]{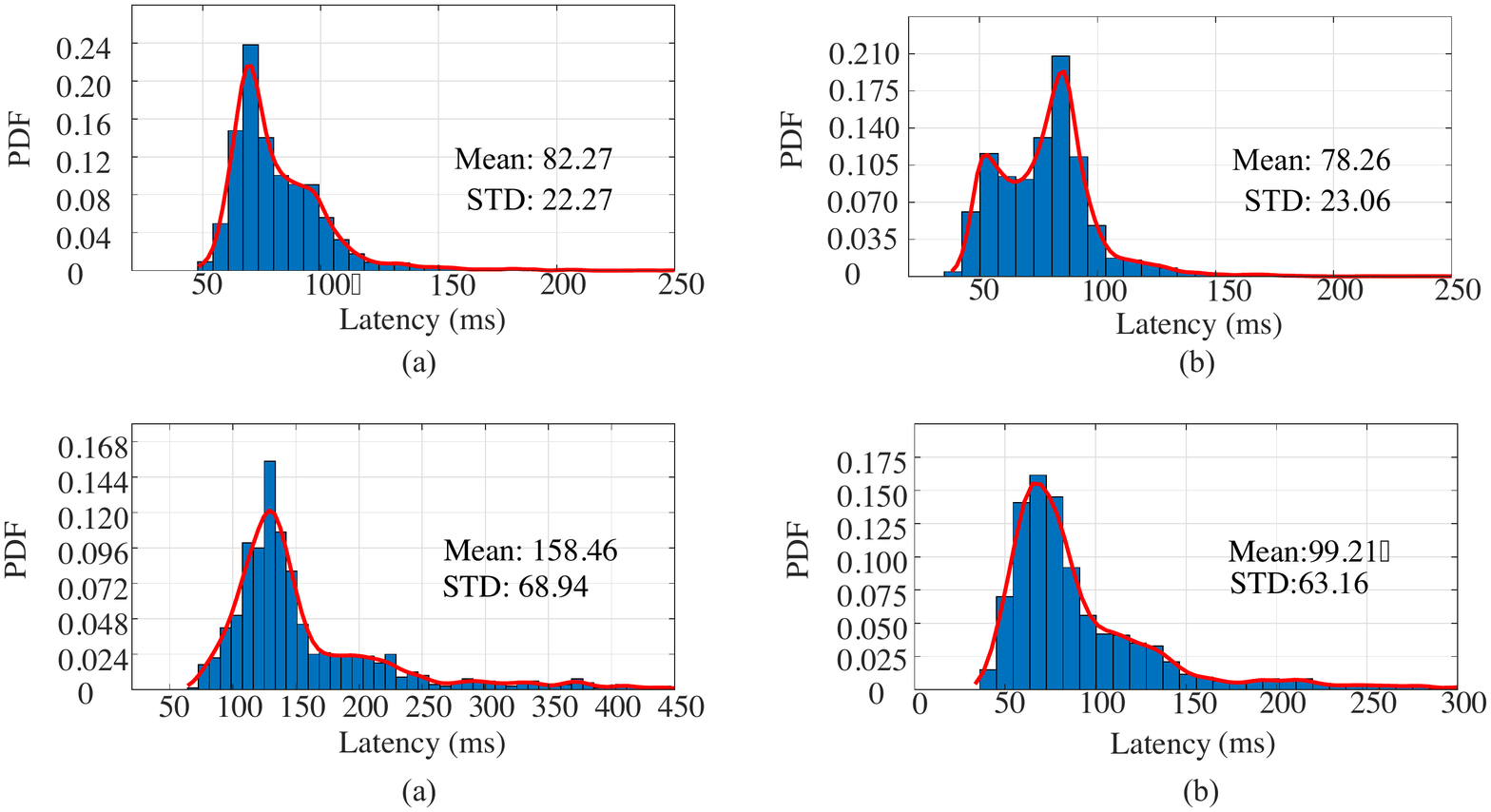}
\vspace{-0.15in}
\caption{\small PDFs of MNO 2's  (a) cloud and (b) fog latencies in a multistory parking lot.} 
\label{Figure_MNO2PDFParkingLot}
\end{figure}



\subsection{Cloud vs. Fog Latency}

\noindent
{\bf Latency and Reliability Tradeoff:}
We present the histogram 
as well as the empirical PDF for cloud and fog latencies when measured at a fixed location (university lab) in Figure \ref{Figure_PDFCloudLatencyFixLoc}. It can be observed that the PDF of fog latency is dual modal, with the first and second peaks occurring at around 54 ms and 87 ms. The 33 ms difference between the two peaks is mainly caused by SR retransmission periodicity (around 20 to 40 ms) and HARQ retransmission delay (around 1 to 8 ms). In \cite{Hadzic2017MECePC}, the authors observed a sawtooth RTT pattern  caused by the SR retransmission periodicity at 20 ms with around 40 ms amplitude in a fixed lab location. Because our latency traces are recorded every 500 ms, we did not observe a clear sawtooth pattern. However, the SR retransmission still contributes to the second peak of the latency traces. as shown in Figure \ref{Figure_PDFCloudLatencyFixLoc}(a). 
From Figures \ref{Figure_PDFCloudLatencyFixLoc}(a) and \ref{Figure_PDFCloudLatencyFixLoc}(b), we can observe that the Internet connection between the LTE network and the cloud server contributes to approximately 12 ms of extra delay in the overall RTT. 
It is interesting to observe that for most of the collected traces, the STD of the cloud latency is less than that of the fog latency.
This means that the extra delay and Internet connection variability somehow compensate for the latency variation of the wireless link between UE and ePC. 
The above observation also verifies the recent study reported in \cite{Inaltekin2018Fog}, where the authors suggest that although the cloud server normally has higher average latency compared to the fog node, the service latency between the UE and cloud exhibits less uncertainty, i.e., lower STD, compared to that between UE and fog node.

In Figures \ref{Figure_MNO1PDFLatencyDriving} and \ref{Figure_MNO2PDFLatencyDriving}, we compare the empirical PDF for cloud and fog latencies for traces collected while driving within two MNO networks. We observe that the mobility of the UE contributes, on average, 10 to 20 ms of extra latency, compared to a fixed location. As indicated in Table \ref{Table_LatencyPerformance}, the driving latency traces exhibit a significantly higher RTT variations, e.g., around 30 ms to 40 ms increase in the 90 percentile of the confidence level for fog and cloud latencies. This can be caused by various factors, such as handover, data loss, and reconnection. We will further discuss this issue later in this section.

\noindent
{\bf Cloud/Fog Server Selection and Adaptation:}
In Figure \ref{Figure_KRDistanceFixedandDrivingComp}, we present the CDFs for fog and cloud latencies and compare their K-R distance under three latency thresholds of 50 ms, 100 ms, and 150 ms, with the weighting factor set to 1. We can observe that for fixed-location traces, the minimum K-R distance between cloud and fog latency occurs at around 85 ms, at which point the difference between fog and cloud latency confidence is only 0.23\%. In other words, if the UE is satisfied with the service quality as long as the latency is below 85 ms, then offloading the workload to either cloud or fog node will not have any noticeable difference in performance. However, for applications that are sensitive to latency below 85 ms, the fog node offers much better performance than the CDC. For example, if the maximum tolerable latency of the UE is 63 ms, the difference between the confidence levels offered by cloud and fog node can be as high as 58.6\%. 

For latency traces collected while driving (Figure \ref{Figure_KRDistanceFixedandDrivingComp} (b)), the average K-R distance between cloud and fog latencies smaller than that of the fixed location. In particular, the minimum K-R distance occurs at 74 ms with only 0.55\% difference between the confidence levels of cloud and fog latencies. The maximum K-R distance is  observed at 101 ms, where switching from a cloud server to a fog server can result in over 16.5\% increase in the confidence level. This  means that the delay uncertainty due to the wireless connection plays a much more dominant role in a driving scenario, compared to the fixed location scenario. 

Fog servers typically have much less computing power compared to cloud servers. For this reason, most existing works suggest offloading only latency-sensitive applications to fog nodes, and leave the more delay-tolerant service workload to the CDC. 
The K-R distance offers a more accurate decision threshold for identifying the services that should be offloaded to CDC or fog nodes. In particular, for a given LTE network at a maximum tolerable delay $r_i$, we can write a simple threshold-based policy for selecting fog or cloud server to process a request of service type $i$:  
\begin{eqnarray}
j &=& \left\{ {\begin{array}{*{20}{l}}
{\{\mbox{Cloud Server}\},}& { \mbox{if } {K(G_i, G'_i)} \le \theta_f } \\
{\{\mbox{Fog Server}\}}& \mbox{ otherwise} 
\end{array}} \right.
\label{eq_CloudFogDecisionPolicy}
\end{eqnarray}
where $G_i$ and $G'_i$ are the empirical CDFs for the cloud and fog latencies at tolerance level $r_i$. $\theta_f$ is the threshold that specifies the  difference between tolerable confidence levels of fog and cloud latencies that can be considered to be negligible for service type $i$.





\begin{figure}
\centering
\includegraphics[width=3.5 in]{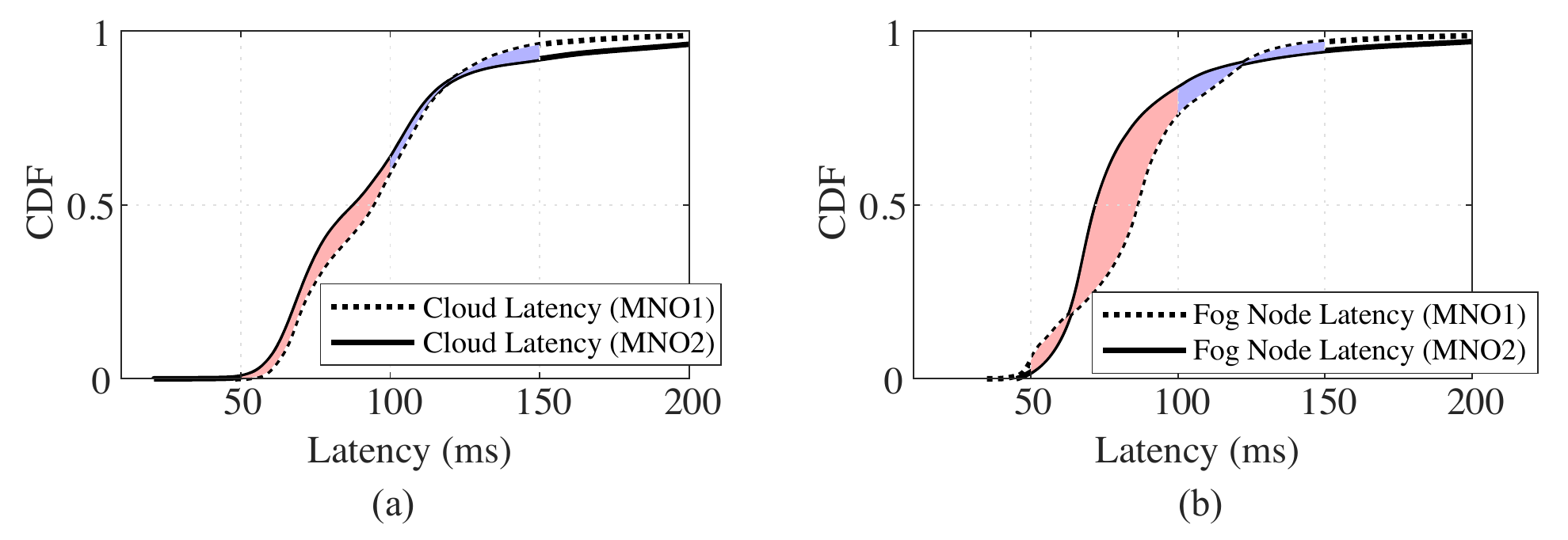}
\vspace{-0.1in}
\caption{\small CDFs and K-R distance of (a) cloud and (b) fog latencies for the two MNOs.} 
\label{Figure_KRDistanceDrivingCloudBTWMNOs}
\end{figure}


\begin{figure}
\centering
\includegraphics[width=3.5 in]{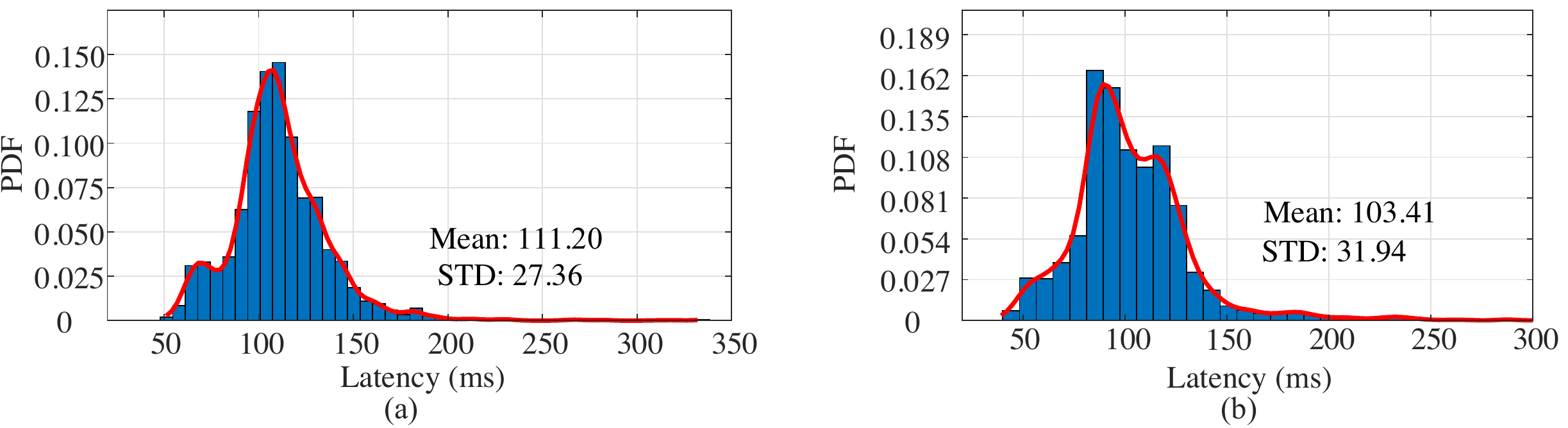}
\vspace{-0.1in}
\caption{\small PDF of MNO 1's cloud and fog latencies during handover between two eNBs.} 
\label{Figure_CloudHandoverATT}
\end{figure}



\subsection{Different MNOs}
The latency experienced under different MNOs exhibits significant spatial variations, depending on the locations and density of deployed eNBs. To investigate the factors that contributed to differences in  latency performances of different MNO networks, we need to look into  specific regions. Figures \ref{Figure_MNO1PDFParkingLot} and \ref{Figure_MNO2PDFParkingLot} depict  the empirical PDF of the measured RTT  in the first level of a multi-story parking garage for two MNO networks. 
For MNO 1, the mean of the measured RTTs increases by around 10ms compared to the mean RTT for the lab location previously presented in Figure \ref{Figure_PDFCloudLatencyFixLoc}. This can be caused by the higher chances of HARQ retransmission and in-synchronization. The RTTs observed over MNO 2's network suffer from a much higher increase in both the average latency as well as its STD. This can be attributed to a less dense deployment of eNBs in the that area compared to MNO 1. Another reason causing the performance degradation for MNO 2  is that its LTE network in that area operates at 1900 MHz. MNO 1, on the other hand, operates at a lower frequency band (850 MHz), where the signal is more capable of penetrating through concrete walls. This will also increase the chances of packet loss, in-synchronization, connection drop, and retransmission.

In Figure \ref{Figure_KRDistanceDrivingCloudBTWMNOs}, 
we compare the CDFs of cloud and fog latencies offered by the two MNOs. For the fog latency, we observe that if the UE latency constraint is 88 ms, then the difference in the confidence levels offered by the two MNOs has a maximum value of 25.79\%. MNO 2 offers a higher confidence level for services with the maximum tolerable latency below 131 ms. The fog latency offered by two MNOs provide the same confidence level at 64 ms and 125 ms. The maximum difference between the fog latency confidence level is at 80 ms. In this case, MNO 2 offers   29.91\% higher confidence than MNO 1.

\subsection{Handover}


Next, we investigate the impact of handover on latency performance. Figure \ref{Figure_CloudHandoverATT} depicts the empirical PDF of the RTT when the UE is driving between two eNBs in a open straight route outside of the city center. Our measurement setup does not allow us to identify the exact location/timing of the handover, i.e., handover can even happen after the UE drives past the target eNB. Therefore, the actual latency performance during handover  will in practice be much worse than the results presented in Figure \ref{Figure_CloudHandoverATT}. Also here we consider the two eNBs located outside of the city center in an open road. The handover process in an urban setting is expected to have much higher latency. Even with these limitations, we can still observe that the average latency for both cloud and fog nodes increases by 40 ms. 

\subsection{Driving Speed}

\begin{figure}%
\center
\includegraphics[width=2.5 in]{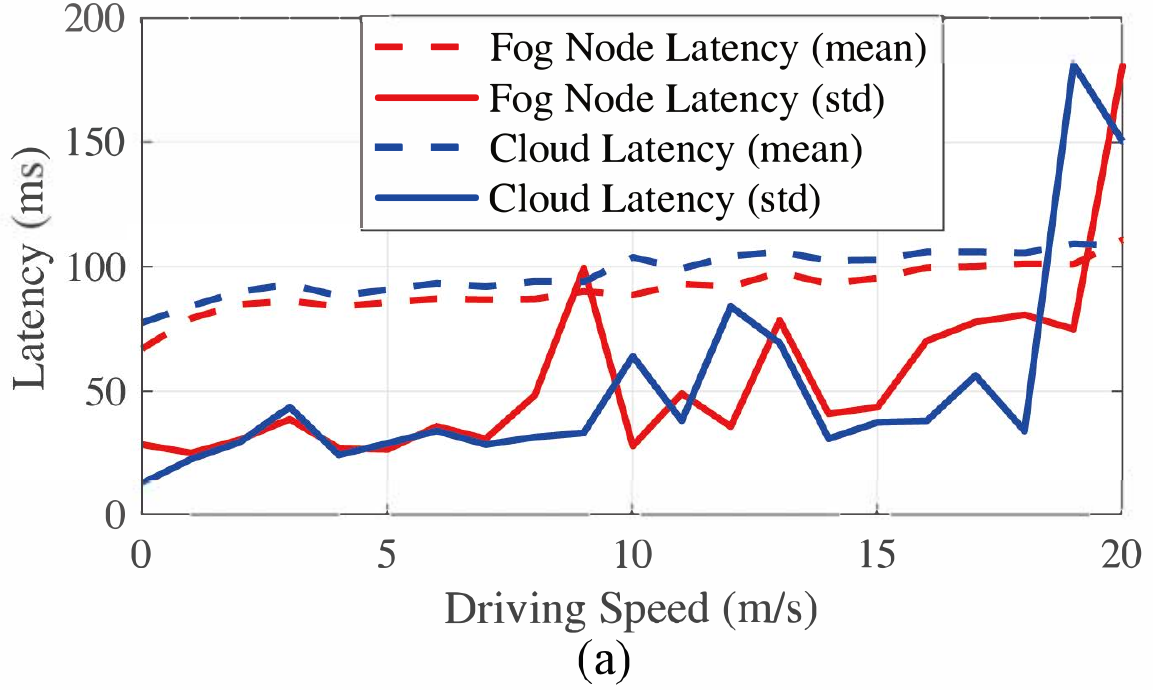}
\includegraphics[width=2.5 in]{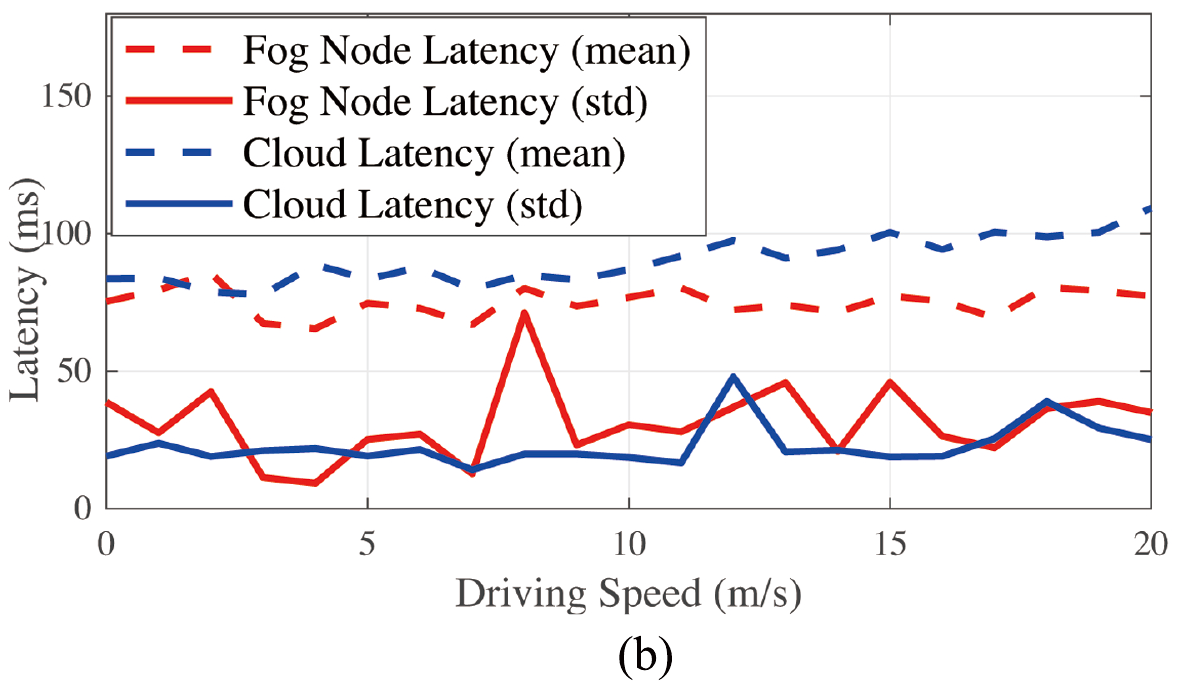}
\caption{Latency vs. driving speed for of (a) traces collected in an small urban region, and (b) all the driving traces.}
\label{Figure_LatencyvsSpeedTotal}
\vspace{-0.1in}
\end{figure}

UE mobility gives rise to Doppler effect, which increases the chances of packet drop and connection failure. To investigate the impact of mobility on service latency in practical system, in Figure \ref{Figure_LatencyvsSpeedTotal}, we analyze the latency traces at different driving speeds. We first present the mean and STD of the latency traces for all datasets collect from our driving measurements. Surprisingly, we did not observe a significant increase in the RTT when the driving speed increases. For example, the average fog latency remains almost the same even when the driving speed approaches 20 m/s. Cloud latency increases by almost 20 ms, on average, when speed reaches 20 m/s. This is because such a speed is possible only when the vehicle is driven outside of the city center. The increased in-synchronization and disconnection probability are compensated for the decrease in blockage experienced inside the city center. In Figure \ref{Figure_LatencyvsSpeedTotal} (a), we compare the latency traces collected within an urban area with high eNB deployment density. Again, we observe a slight increase in the average cloud and fog latencies, i.e., around 10-20 ms for both. However, we also observe a significant increase in the STD of both latencies when the speed exceeds 10 m/s.

\section{Optimal Network/Server Selection and Adaptation}
\label{Section_OptimalPolicy}
In this section, we derive the conditions and optimal policy for any UE to dynamically switch between different wireless access networks so as to maximize its service latency confidence while driving. 
%
%
\blu{
As mentioned before, in addition to network characteristics, UE latency also depends on the driving pattern, which is affected by the driver's intended destinations (e.g., home and office locations), driving habits, time of day, and traffic conditions. These factors further accentuate the complexity of determining the optimal policy. Nonetheless, previous literature as well as our own measurements suggest that the temporal and spatial latency confidence exhibits the following features:
%
\begin{itemize}
\item[(1)] Vehicle movements, including location and driving speed, are approximately Markovian, i.e., the vehicle's future location and speed mainly depend on the current location and speed.
%
\item[(2)] The K-R distance of the latency confidence for two different wireless access networks can be assumed to be fixed at any given location.
\item[(3)] The forecasting window used in the optimal decision-making process will also impact the UE policy for switching between different MNOs. In particular, if a vehicle tries to optimize the expected latency over a fixed number of future time slots, i.e., while driving through a selected route from a given source to a destination, the network selection/switching problem can be modeled as a finite-horizon problem in which a sequence of network switching decisions should be made during the entire route. Alternatively, the vehicle may try to maximize its long-term (infinite-horizon) expected latency confidence. In this case, it is more reasonable for the vehicle to seek a long-term policy that specifies the ideal network switching condition under each possible environmental and driving state.
\end{itemize}
}

Motivated by the above observations, we formulate the network adaptation and fog/cloud server selection at the UE as a Markov decision process (MDP), in which the UE tries to find the optimal switching policy while driving. More formally, we consider an MDP ${\cal P} = \langle {\cal S}, {\cal A}, {\Gamma}, U \rangle$ that consists of the following elements: 


\noindent
{\bf State Space ${\cal S}$:} The state includes the UE's driving speed $v$, its location $x$, and the connected wireless access network $l$. Typically, $v$ and $x$ are continuous variables. However, based on  our previous discussion, real-world latency performance does not seem to be sensitive to small changes in these two parameters. Furthermore, the empirical PDFs for $v$ and $x$ were obtained using only a finite number of 
states.
We can therefore define the state space $\cal S$ as a finite set of possible intervals of speed, location regions, and LTE network choices. We write each instance of state as $s = \langle v, x, l \rangle$ for $s \in {\cal S}$.

\noindent
{\bf Action Space ${\cal A}$:} Suppose the UE is connected to MNO $l$. It can then decide whether or not to switch to another MNO $k$, $k\neq l$, or stay with the current choice. We assume a service request  can be submitted to only one LTE network at a time. We define the action of the UE as a binary 1/0 choice, indicating whether or not to switch to MNO $k$. 
We write the action set as ${\cal A} = \{ 0, 1 \}$.

\noindent
{\bf State Transition Function $\Gamma$:} The probability of transiting from one possible location and driving speed to another location and speed can be estimated from our measurement dataset.
We observe that the driving speed as well as its probability of transiting to another possible speed is closely related to the driving time, i.e., they differ between peak and non-peak hours.  
Therefore, we consider different state transition probabilities at different time slots throughout a day. To simplify our description, we assume the state transition probability can be considered as fixed during the time of driving, and write the probability of transitioning from state $s$ to $s'$ when taking action $a$ as $\Gamma \left( s', s, a \right) = \Pr \left( s' | s, a \right)$.

\noindent
{\bf Utility Function $U$:} Our main objective is to maximize the UE confidence level in 
successfully delivering all requested services within the required latency. We assume the UE can receive requests from a set of services ${\cal M}$. In each time slot, a service request of type $i$, $i \in {\cal M}$, is generated with service request probability 
$p_i$. Let $r_i$ be the maximum tolerable delay for type $i$ service. To avoid switching  back-and-forth between different MNOs, we assume a fixed cost $c_t$ for switching from one MNO to another. We consider the instantaneous utility function $u_t$ defined in (\ref{eq_InstantUtility}).  


\begin{theorem}
\label{Theorem_ThresholdPolicyFinite}
For 
a vehicular system with two available MNOs, 
the optimal policy for the UE to decide whether or not to switch from one MNO to another is a threshold policy (e.g. the policy is in the form of (\ref{eq_CloudFogDecisionPolicy})). In particular, suppose a UE is connected to an MNO with empirical latency CDF $F_t$ at time slot $t$. Let $G_t$ be the empirical latency CDF of the other MNO.
\blu{
\begin{itemize}
\item[(1)] For finite-horizon decision making, the optimal UE policy  is given by:
%
\begin{eqnarray}
%
\pi^* \left(s_t\right) = \left\{ {\begin{array}{*{20}{l}}
{1,}& { \mbox{if } {K(F_t, G_t) \le {\Delta_{t}-c_t}}} \\
{0,}& \mbox{otherwise} 
\end{array}} \right.
\label{eq_ThresholdPolicyFinite}
\end{eqnarray}
where ${\Delta_{t}}$ $=$ $\sum_{\langle v_{t+1}, x_{t+1} \rangle \in {\cal V}\times{\cal X}}$ $\Pr\left( \langle v_{t+1}, x_{t+1} \rangle|\langle v_{t}, x_{t} \rangle \right)$ $Y \left( v_{t+1}, x_{t+1} \right)$ and
\begin{eqnarray}
\lefteqn{ Y \left( v_{t+1}, x_{t+1} \right)} \label{eq_DeltatFinite}\\ 
&=& \left\{ {\begin{array}{*{20}{l}}
{c_t,}& { \mbox{if } {K(F_t, G_t) < {\Delta_{t+1}-c_t}}} \\
{-c_t,}& { \mbox{if } {K(F_t, G_t) > {\Delta_{t+1}+c_t}}} \\
{\Delta_{t+1}-K(F_t, G_t),}& \mbox{otherwise}. 
\end{array}} \right.\nonumber
\end{eqnarray}
\item[(2)] For infinite-horizon decision making, the optimal UE policy is given by:
\begin{eqnarray}
%
{\pi'}^* \left(s\right) = \left\{ {\begin{array}{*{20}{l}}
{1,}& { \mbox{if } {K(F_t, G_t) \le {\gamma \Delta - c_t }}} \\
{0}& \mbox{otherwise} 
\end{array}} \right.
\label{eq_ThresholdPolicyInFinite}
\end{eqnarray}
where ${\Delta} = \sum_{\langle v', x' \rangle \in {\cal V}\times{\cal X}} \Pr\left( \langle v', x' \rangle|\langle v, x \rangle \right)Y \left( v', x' \right)$ and
\begin{eqnarray}
\lefteqn{ Y \left( v', x' \right)} \label{eq_DeltatInFinite} \\ 
&=& \left\{ {\begin{array}{*{20}{l}}
{c_t,}& { \mbox{if } {K(F_t, G_t) < {\gamma\Delta - c_t}}} \\
{-c_t,}& { \mbox{if } {K(F_t, G_t) > {\gamma\Delta + c}}} \\
{-K(F_t, G_t)+\gamma\Delta,}& \mbox{otherwise} 
\end{array}} \right. \nonumber
\end{eqnarray}
\end{itemize}
}
\end{theorem}
\begin{IEEEproof}
Theorem \ref{Theorem_ThresholdPolicyFinite} follows directly from the standard policy iteration.
We first consider the finite-horizon scenario. To maximize the utility sum for a given number of time slots, the UE must consider both the current utility and the expected future utility obtained during the remaining time slots, i.e., the UE maximizes the following function:
\begin{eqnarray}
\lefteqn{V_t \left( a_t, s_t \right) =} \nonumber \\
&& \left( 1 - a_t \right) \left[ \sum_{i\in {\cal M}} w_i F_{i,t} \left( s_t, l \right) + \sum_{s_{t+1}\in {\cal S}} \Gamma\left( s_{t+1}, s_t, l \right) \right. \nonumber\\
&&\;\;\;\; \left.  V^*\left( s_{t+1}, a^*_{t+1} \right) \right] + a_t \left[ \sum_{i\in {\cal M}} w_i G_{i,t} \left( s_t, k \right) + \right. \nonumber \\
&& \left. \sum_{s_{t+1}\in {\cal S}} \Gamma\left( s'_{t+1}, s_t, k \right) V^*\left( s'_{t+1}, {a'}^*_{t+1} \right) - c_t \right].
\end{eqnarray}

We can therefore write the optimal policy as
\begin{eqnarray}
\lefteqn{ \pi^* \left(s_t\right) = \arg \max\limits_{a\in \{0, 1\}} V_t \left( a, s_t\right)} \nonumber \\
&=& \arg \max\limits_{a\in \{0, 1\}} a\left[ w_i \sum_{i\in {\cal M}} \left( G_{i,t} \left( s_t, k \right) - F_{i,t} \left( s_t, l \right) \right) - c_t \right. \nonumber \\
&& \left. + \sum_{s'_{t+1}\in {\cal S}} \Gamma\left( s'_{t+1}, s'_t, k \right) V^*\left( s'_{t+1}, {a'}^*_{t+1} \right) \right. \nonumber \\
&& \left. - \sum_{s_{t+1}\in {\cal S}} \Gamma\left( s_{t+1}, s_t, l \right) V^*\left( s_{t+1}, {a}^*_{t+1} \right) \right] \nonumber \\
&=& \arg \max\limits_{a\in \{0, 1\}} a\left[ -K \left( F_t, G_t \right) - c_t  + \Delta_{t} \right] \nonumber \\
&=& \left\{ {\begin{array}{*{20}{l}}
{1,}& { \mbox{If } {K(F_t, G_t) \le { \Delta_{t}-c_t}}}, \\
{0}& \mbox{Otherwise}, 
\end{array}} \right.
\end{eqnarray}
where $\Delta_{t}$ is given by 
\begin{eqnarray}
\lefteqn{ \Delta_{t} = \sum_{s'_{t+1}\in {\cal S}} \Gamma\left( s'_{t+1}, s_t, k \right) V^*\left( s'_{t+1}, {a'}^*_{t+1} \right)} \nonumber \\
&&\;\;\;{- \sum_{s_{t+1}\in {\cal S}} \Gamma\left( s_{t+1}, s_t, l \right) V^*\left( s_{t+1}, {a}^*_{t+1} \right) } \nonumber\\
&&= \sum_{s_{t+1}\in {\cal S}} \Pr\left( \langle v_{t+1}, x_{t+1} \rangle | \langle v_{t}, x_{t} \rangle \right)\cdot \nonumber \\
&&\;\;\;\;\;\;\; \left[ V^*\left( \langle v_{t+1}, x_{t+1}, k \rangle, {a'}^*_{t+1} \right) \right. \nonumber \\
&&\;\;\;\;\;\;\;\;\;\;\;\;\;\;\;\;\;\;\;\;\; \left.- V^*\left( \langle v_{t+1}, x_{t+1}, l \rangle, {a}^*_{t+1} \right) \right]
\label{eq_Proof_Deltat}
\end{eqnarray}

By substituting the optimal policy ${a}^*_{t+1} = \pi^* \left(s_{t+1}\right)$ into (\ref{eq_Proof_Deltat}), we obtain the result in Equation (\ref{eq_DeltatFinite}).
The infinite-horizon decision making scenario follows a similar approach. We omit the details due to the limit of space. 
\end{IEEEproof}


\red{Theorem \ref{Theorem_ThresholdPolicyFinite} offers a simple threshold-based policy for connected vehicles to improve the confidence of latency-sensitive services while driving in highly dynamic environments. More specifically, if the UE wishes to maximize its confidence when driving in a selected route, i.e., for the finite-horizon case, it only needs to keep track of the value of $\Delta_t$, which is mainly affected by its state transition, and compare this value  with the K-R distance between latency performance offered by different networks. If the UE tries to maximize the long-term confidence of wireless latency without any preset termination time, i.e., the infinite-horizon case, it can simply evaluate the possible transition probabilities from the current state, including the current location, speed, and connected network, to the next one and then calculate the value of $\Delta$ to be compared with the K-R distance between different networks for deciding the switching action. }

\red{
The K-R distance between the confidence levels of two wireless access networks can be pre-calculated using previously collected data. It can be stored in a table at the fog node. For example, a fog node 
can utilize crowdsourced data provided by UEs to create and maintain a table of the K-R distances and possible state transitions at different locations, e.g., each fog node will first record the service latencies when serving different UEs at different locations with different driving speeds and network connections and then calculate the empirical PDF as well as the corresponding  K-R distances using the approaches described in Sections \ref{Section_LTELatency} and \ref{Section_MeasureAnalysis}. 
When a UE drives into a new location, the closest fog node performs a table search and comparison, and sends a binary decision variable to the UE, specifying the network switching decision.}

\begin{figure}%
\center
\includegraphics[width=1.8 in]{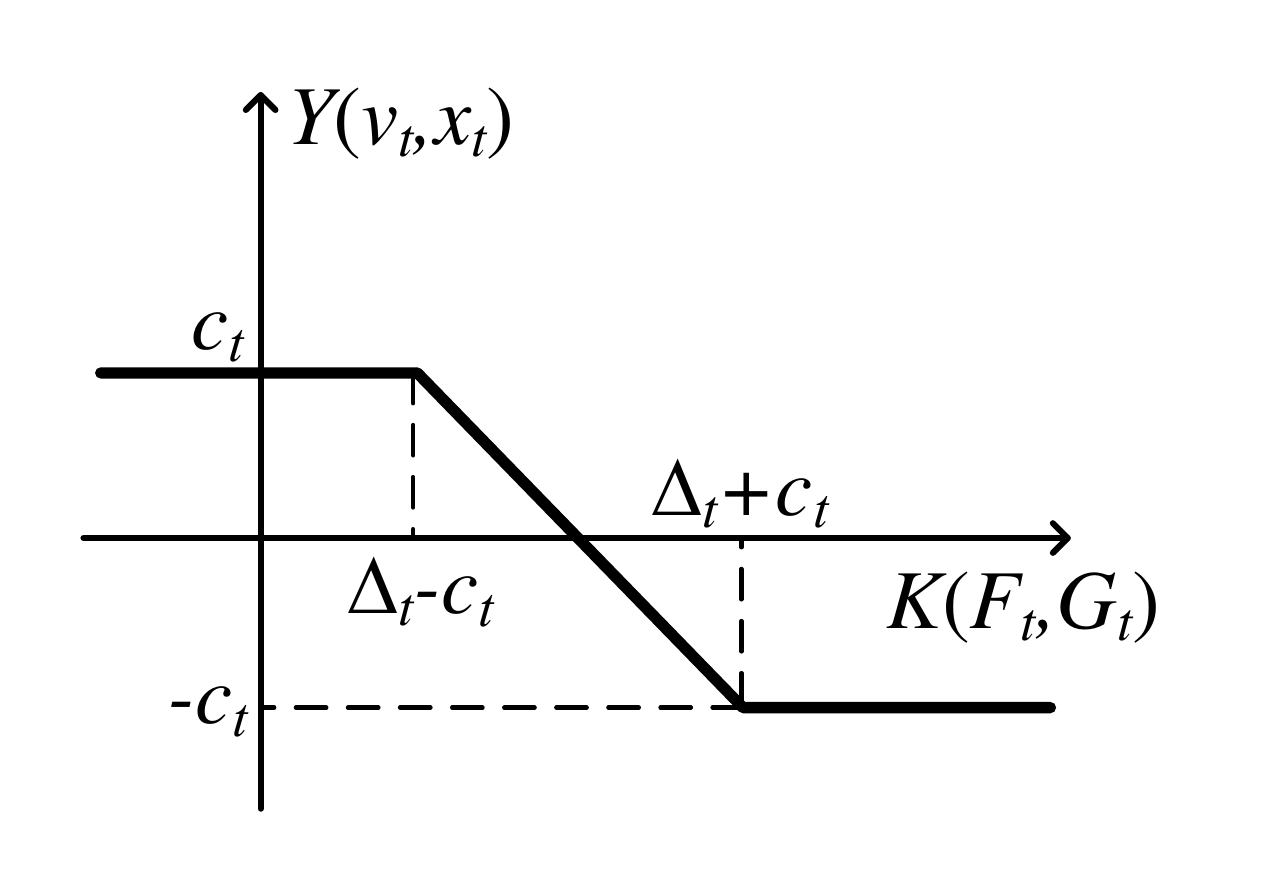}
\vspace{-0.1in}
\caption{Values of $Y \left( v_{t+1}, x_{t+1} \right)$ under different K-R distances $K \left( F_t, G_t \right)$ in time slot $t$.}
\label{Figure_Proof_YvsK}
\vspace{-0.1in}
\end{figure}

Note that $Y \left( v_{t+1}, x_{t+1} \right)$ and $Y \left( v', x' \right)$ are the accumulated current and future utilities for the UE starting from time slot $t+1$ in finite and infinite horizon scenarios, respectively. In Figure \ref{Figure_Proof_YvsK}, we present the values of $Y \left( v_{t+1}, x_{t+1} \right)$ under different K-R distances $K \left( F_t, G_t \right)$ in time slot $t$. We can observe that even if the UE adopts the optimal policy in Theorem \ref{Theorem_ThresholdPolicyFinite}, the accumulated  utility loss due to an incorrect selection of the MNO in a single time slot $t$ will not exceed $c_t+K\left( F_t, G_t \right)$. \blu{In other words, even if the UE chooses the wrong MNO network with lower performance in the future, the total performance loss of the UE will not affect the long-term performance as long as it starts to adopt the optimal policy in the future.}

\blu{
The table search and comparison for policies (\ref{eq_ThresholdPolicyFinite}) and (\ref{eq_ThresholdPolicyInFinite}) can be further simplified when the vehicle is driving to some locations with a large K-R distance between the wireless access networks. This is formalized in the following result. 
\begin{proposition}
\label{PropositionSwitchLargeKRDistance}
For both finite-horizon and infinite-horizon decision making processes, the UE should always switch to the other wireless access network if $K \left( F_t, G_t \right) \le -2 c_t$.
\end{proposition}
}

\blu{
Proposition \ref{PropositionSwitchLargeKRDistance} indicates that if the performance gain that results from switching from the currently connected wireless network to the other network is larger than the cost of switching back-and-forth between these two networks, then the UE should always make the switch without considering the future driving decision. In our measurements, we observe that due to the difference in network infrastructure deployments as well as the traffic demands, many locations, e.g., cell-edge locations, can observe a large K-R distance between the two tested MNOs. 
}

\red{Proposition \ref{PropositionSwitchLargeKRDistance} can be further utilized to simplify the policies proposed in Theorem \ref{Theorem_ThresholdPolicyFinite}. In particular, 
according to Proposition \ref{PropositionSwitchLargeKRDistance}, UEs should always connect to the higher-performance network when driving into locations with high K-R distance. }



\section{Numerical Results}
\begin{figure}
\includegraphics[width=3.5 in]{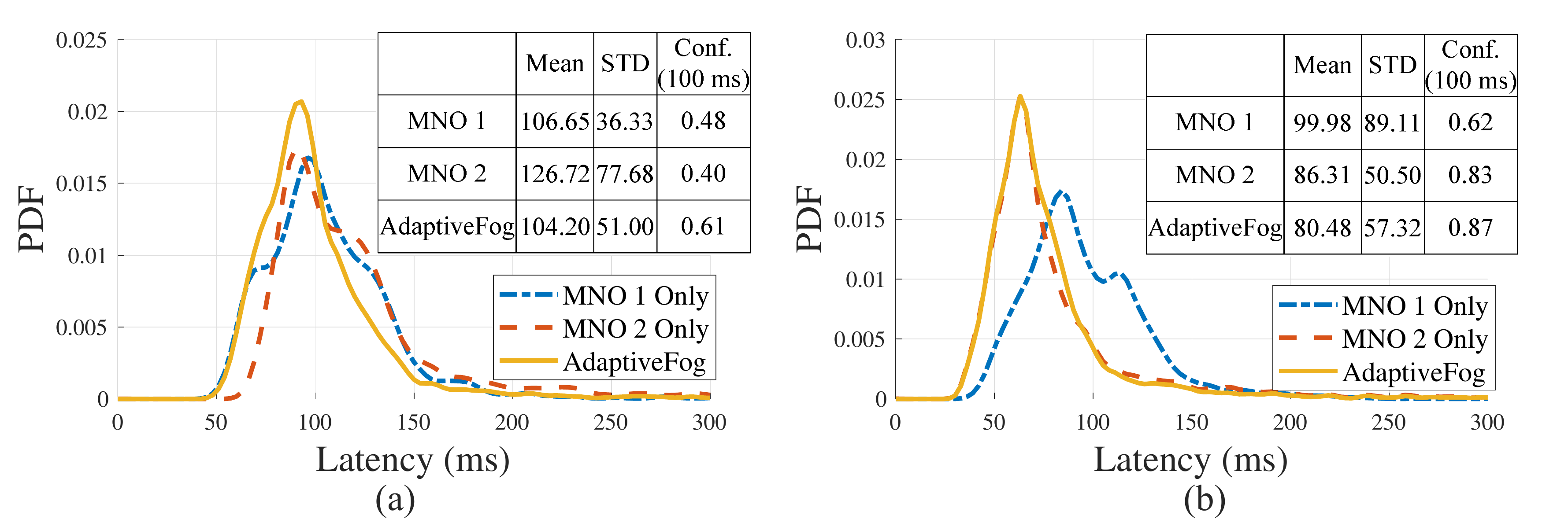}
\vspace{-0.1in}
\caption{\small PDFs of latency for: (a) fog, and (b) cloud using either AdaptiveFog or a single MNO.}
\label{Figure_FogNodeAdaptiveFog}
\end{figure}

%
%
%
%
%
%
%
%
%

\begin{figure}%
\begin{minipage}[t]{0.5\linewidth}
\includegraphics[width=1.8 in]{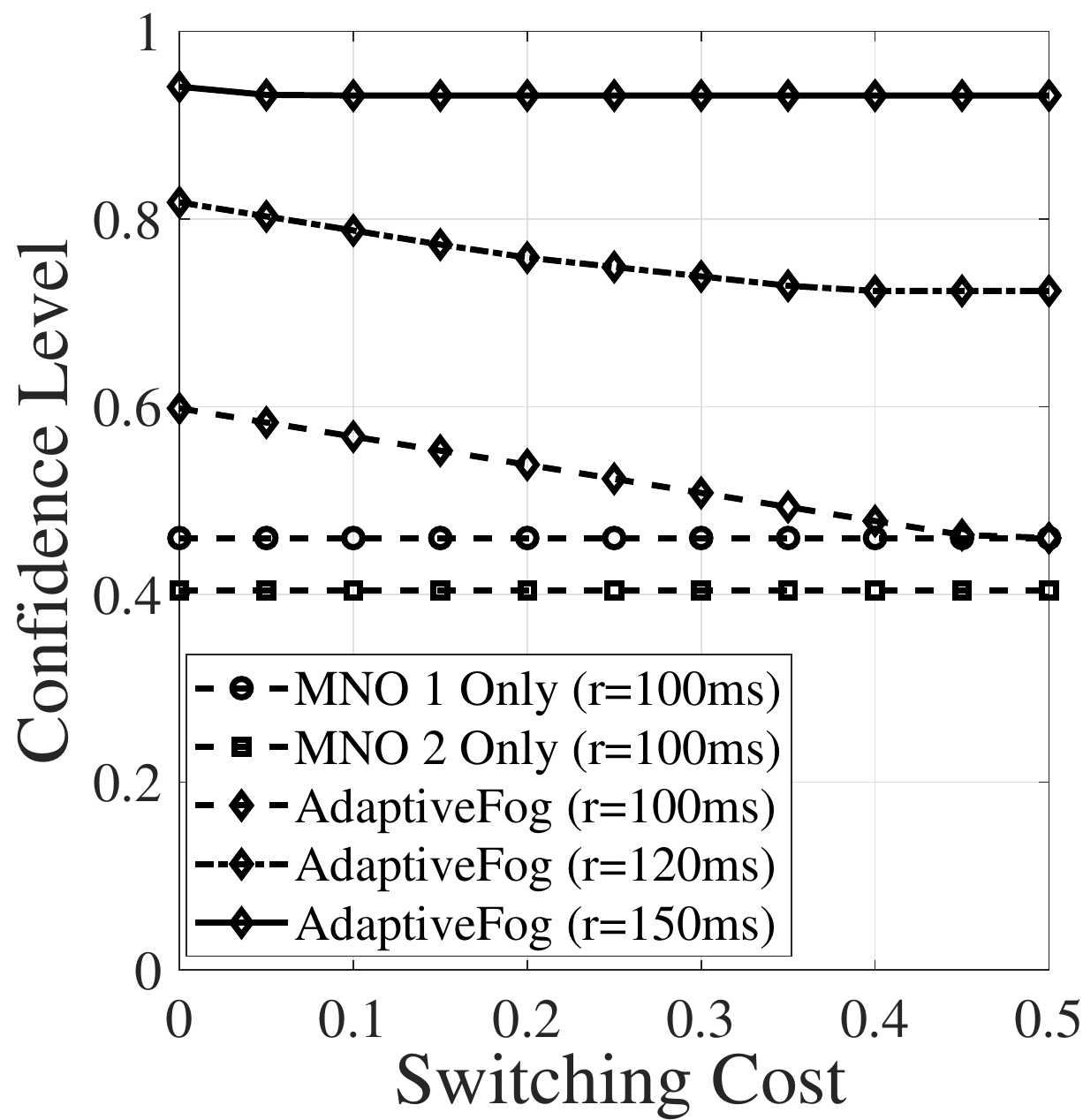}
\vspace{-0.1in}
\captionsetup{labelformat=empty}
\caption*{(a)}
\label{Figure_ConfidenceVSSwitchingCostCloudFinite}
\end{minipage}
\begin{minipage}[t]{0.45\linewidth}
\includegraphics[width=1.8 in]{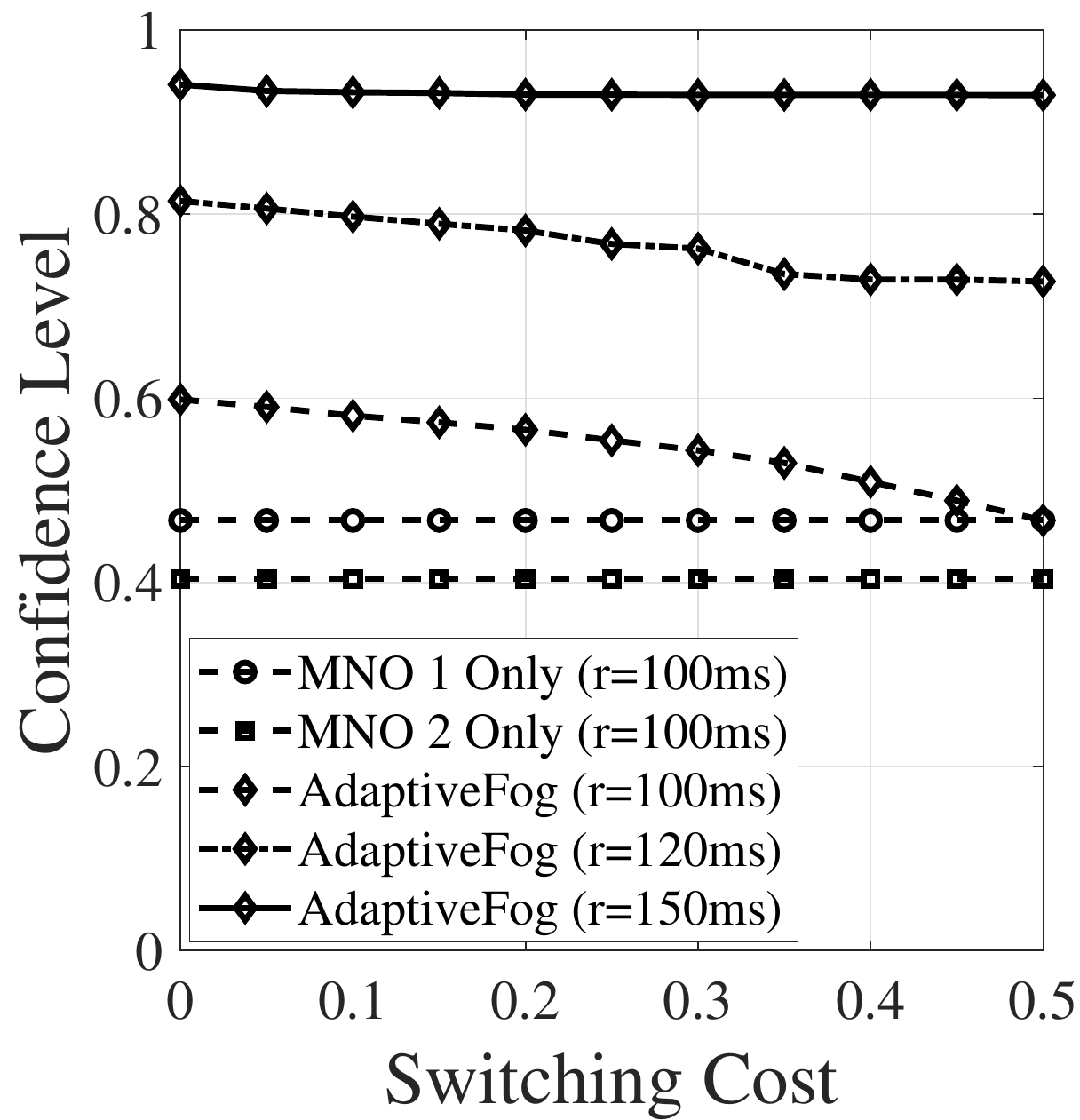}
\vspace{-0.15in}
\captionsetup{labelformat=empty}
\caption*{(b)}
\label{Figure_ConfidenceVSSwitchingCostCloudInfinite}
\end{minipage}
\caption{\small Confidence level for cloud latency under (a) finite- and (b) infinite-horizon decision making.}
\label{Figure_ConfidenceVSSwitchingCostCloudFiniteandInfinite}
\end{figure}

\begin{figure}%
\begin{minipage}[t]{0.5\linewidth}
\includegraphics[width=1.8 in]{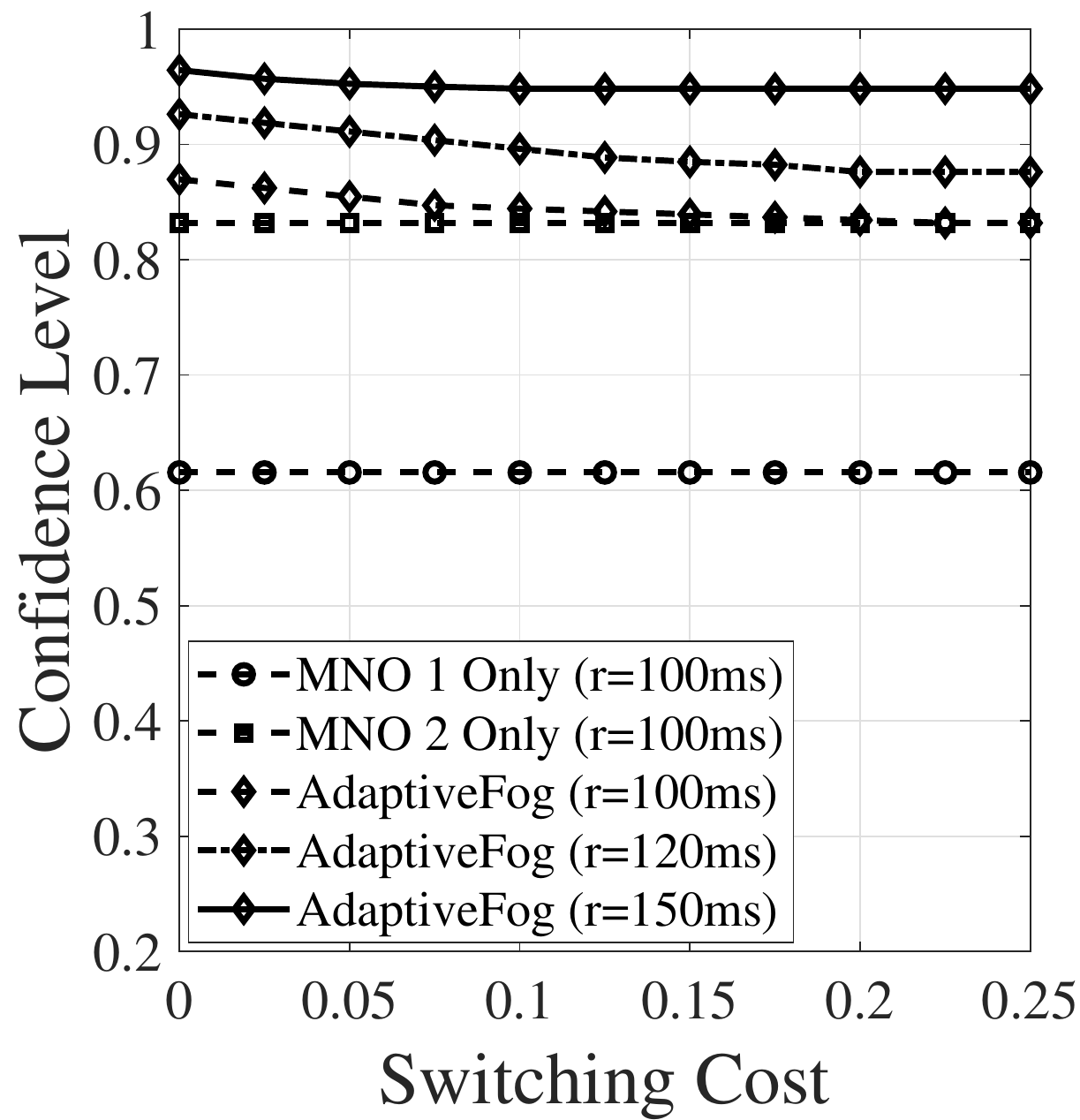}
\captionsetup{labelformat=empty}
\caption*{(a)}
\label{Figure_ConfidenceVSSwitchingCostFogFinite}
\end{minipage}
\begin{minipage}[t]{0.45\linewidth}
\includegraphics[width=1.8 in]{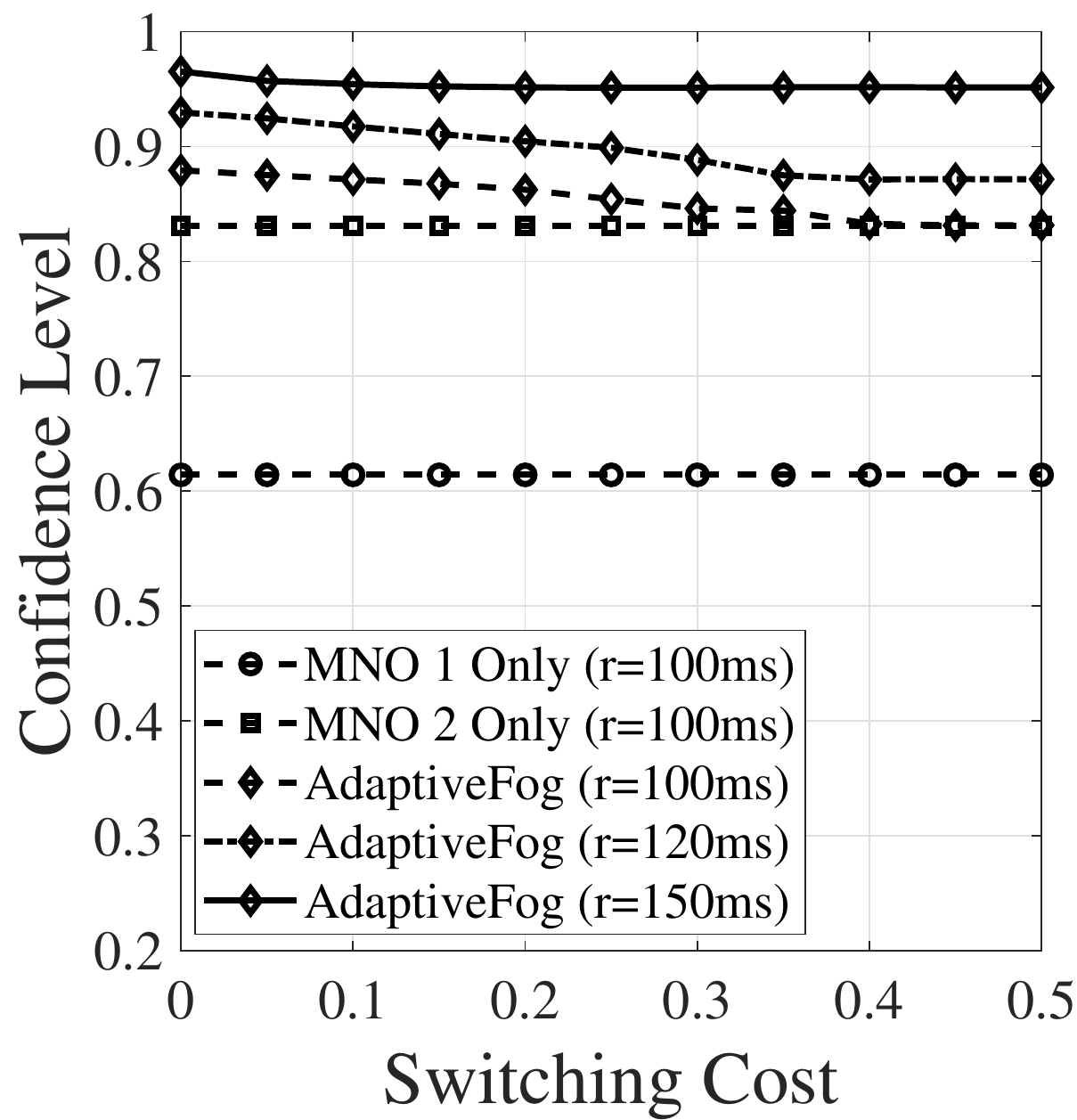}
\captionsetup{labelformat=empty}
\caption*{(b)}
\label{Figure_ConfidenceVSSwitchingCostFogInfinite}
\end{minipage}
\vspace{-0.1in}
\caption{\small  Confidence level of fog latency under (a) finite- and (b) infinite-horizon decision making.}
\label{Figure_ConfidenceVSSwitchingCostFogFiniteandInfinite}
\end{figure}
\begin{figure}%
\begin{minipage}[t]{0.5\linewidth}
\includegraphics[width=1.8 in]{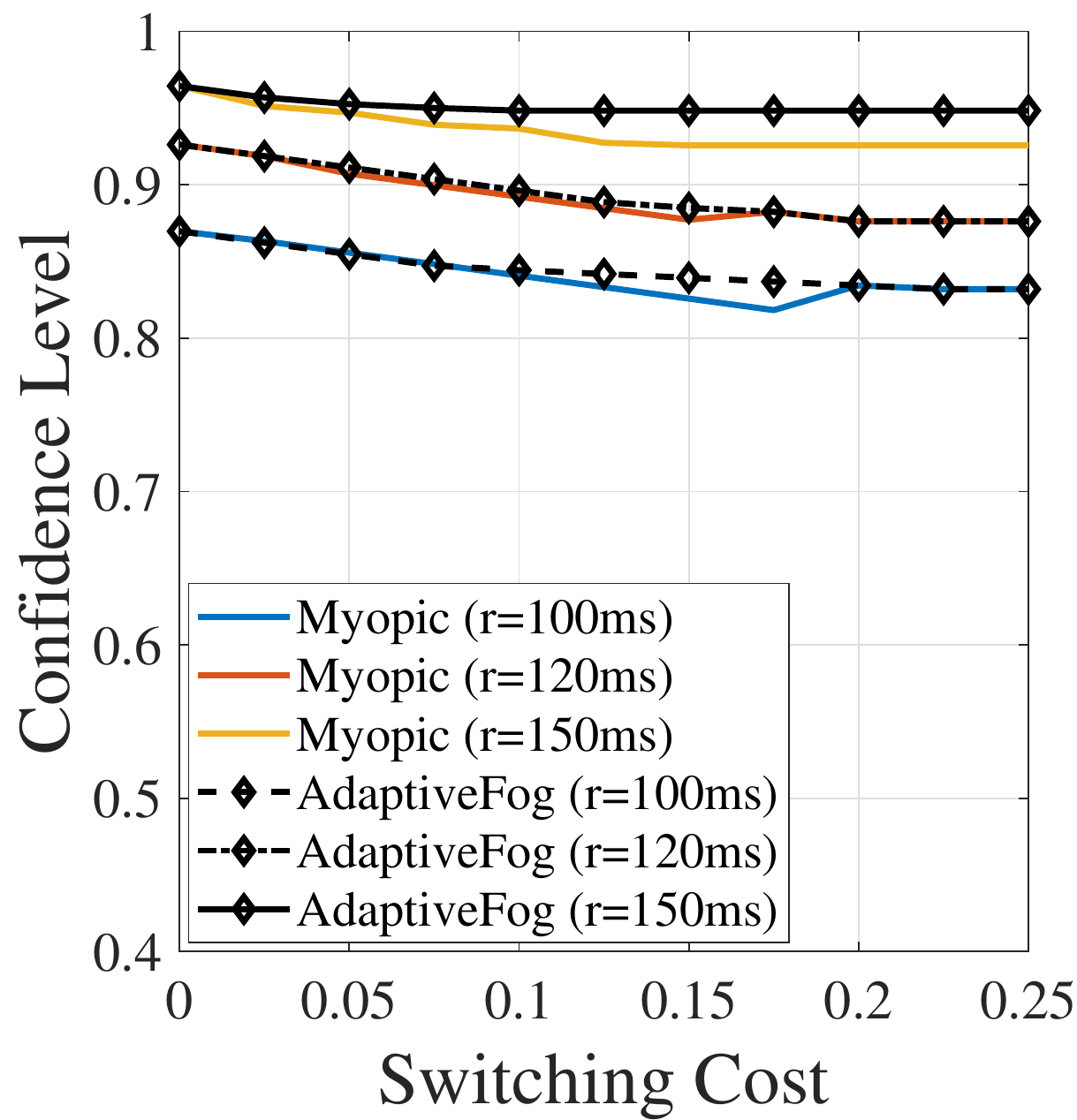}
\captionsetup{labelformat=empty}
\caption*{(a)}
\label{Figure_ConfidenceVSSwitchingCostFogFinite}
\end{minipage}
\begin{minipage}[t]{0.45\linewidth}
\includegraphics[width=1.8 in]{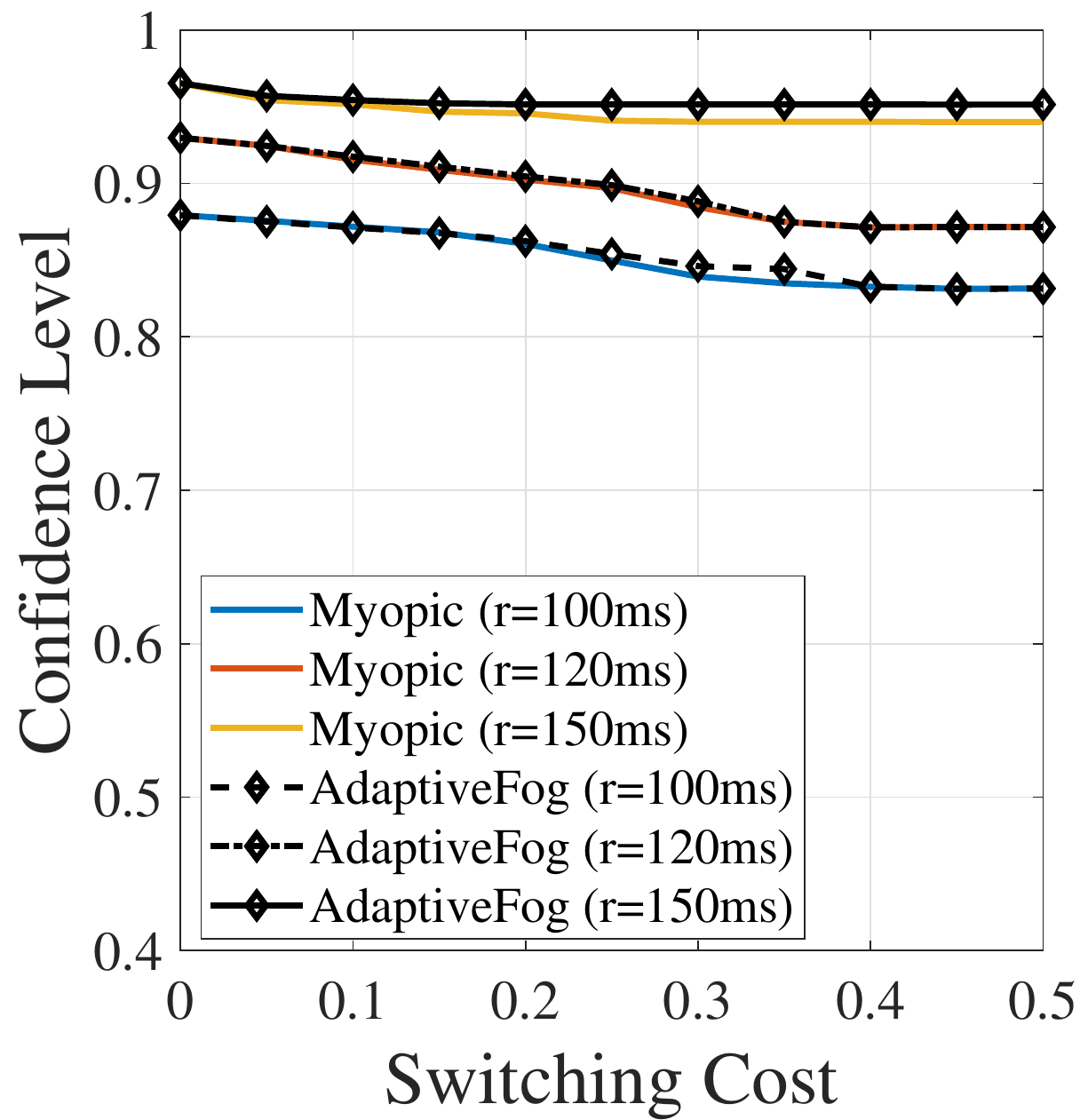}
\captionsetup{labelformat=empty}
\caption*{(b)}
\label{Figure_ConfidenceVSSwitchingCostFogInfinite}
\end{minipage}
\vspace{-0.1in}
\caption{\blu{\small  Confidence level for fog latency under: (a) finite-, and (b) infinite-horizon, compared with myopic strategy.}}
\label{Figure_ConfidenceVSSwitchingCostFogFiniteandInfiniteMyopic}
\end{figure}

\begin{figure}%
\begin{minipage}[t]{0.5\linewidth}
\includegraphics[width=1.8 in]{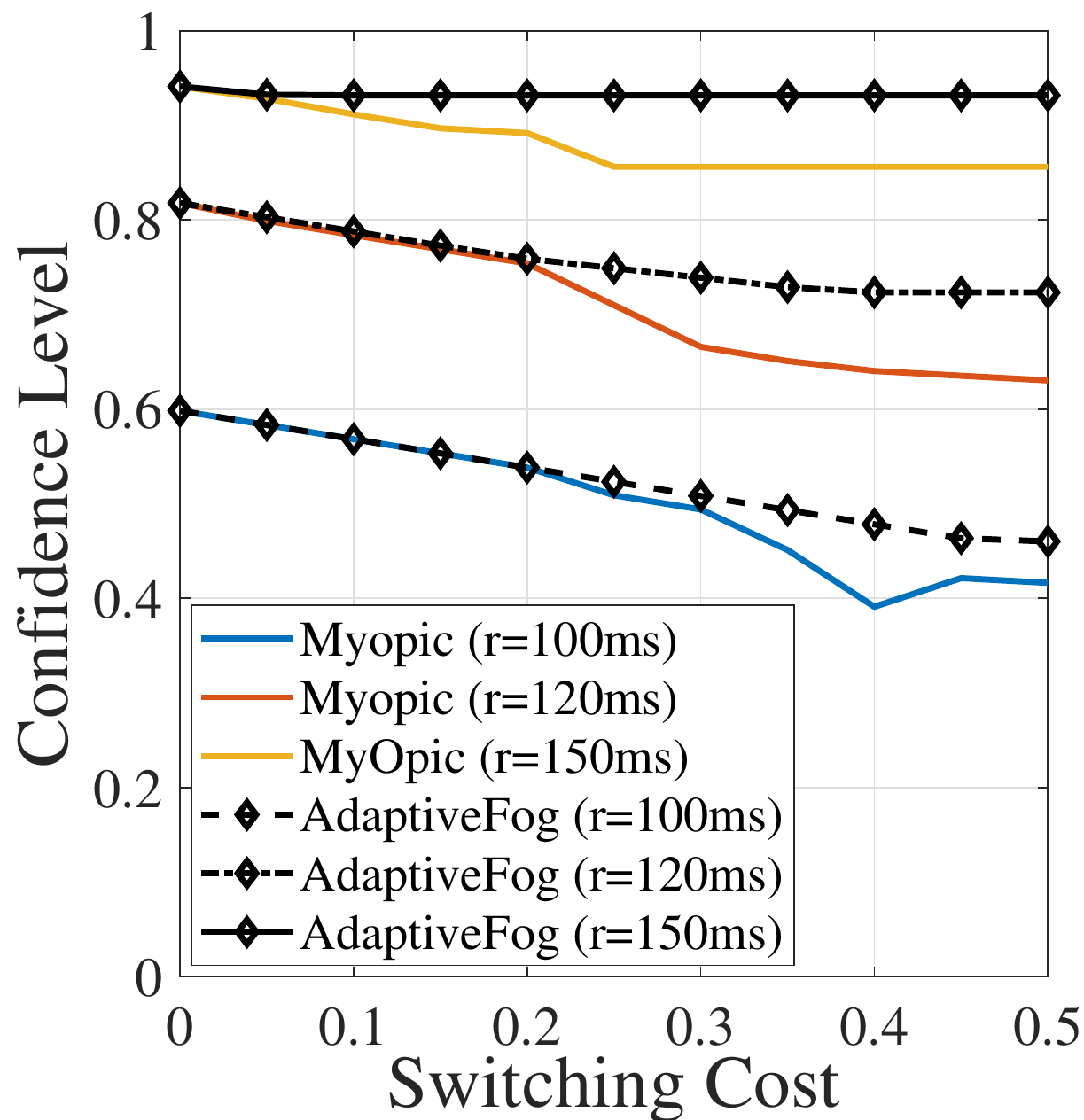}
\captionsetup{labelformat=empty}
\caption*{(a)}
\end{minipage}
\begin{minipage}[t]{0.45\linewidth}
\includegraphics[width=1.8 in]{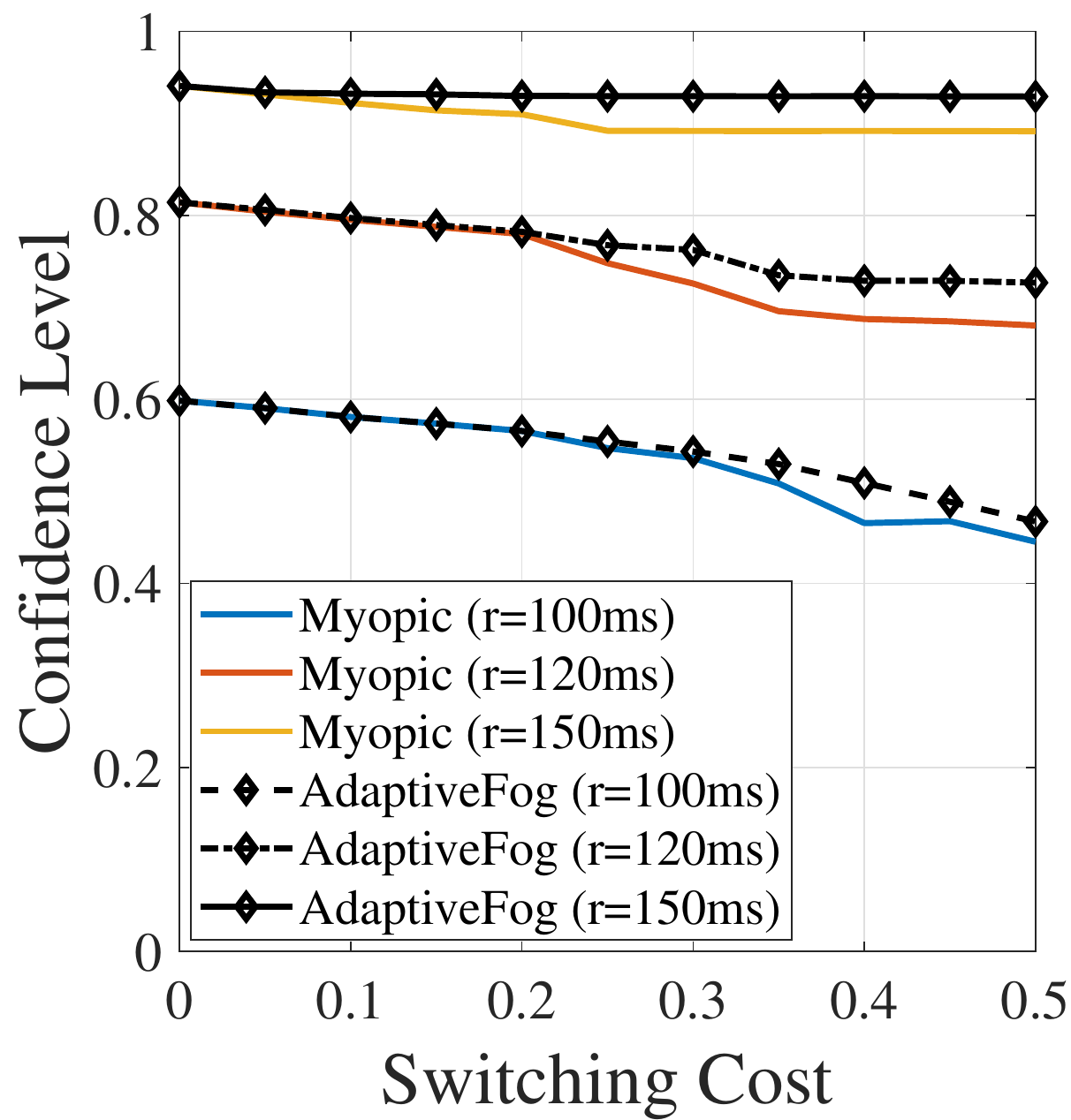}
\captionsetup{labelformat=empty}
\caption*{(b)}
\end{minipage}
\vspace{-0.1in}
\caption{\blu{\small  Confidence level of cloud latency under: (a) finite-, and (b) infinite-horizon, compared with myopic strategy.}}
\label{Figure_ConfidenceVSSwitchingCostCloudFiniteandInfiniteMyopic}
\end{figure}



\blu{
In this section, we evaluate the latency performance of AdaptiveFog using the driving dataset collected in our measurement campaign. We focus on the latency confidence for the cloud and fog latency (i.e., RTT) under different switching costs. We use one week worth of measured data with similar driving routes and time duration captured on a daily basis. The first three days of data are used as the training dataset to calculate the network switching policy. This policy is then tested by applying it for the rest of the week.}

We first evaluate the impact of applying AdaptiveFog on the PDF of the RTT for driving scenario. We compare the results to the single MNO case (no switching). In Figure \ref{Figure_FogNodeAdaptiveFog}, we present the empirical PDFs \blu{as well as the mean, STD, confidence level at 100 ms for both fog and cloud latencies. We can observe that AdaptiveFog  provides significant improvement in fog latency, with around 22 ms and 2 ms reduction in the average latency, compared with MNOs 1 and 2, respectively.   More importantly, AdaptiveFog significantly increases the confidence of the UE by over 20\% compared to the worst performing MNO. For the cloud latency, the average latency improvement are around 19 ms and 6 ms for MNOs 1 and 2, respectively. An  improvement in the confidence level is observed, with 25\% and 4\% gain over MNOs 1 and 2, respectively.
}

It is obvious that the performance of AdaptiveFog strongly depends on the switching cost. \blu{Particularly, if the switching cost is close to zero, the UE should always switch to the MNO network that offers the highest confidence level. If the cost for switching to another network is large, the incentive for the UE to switch to another MNO will be reduced.} In Figures \ref{Figure_ConfidenceVSSwitchingCostCloudFiniteandInfinite}--\ref{Figure_ConfidenceVSSwitchingCostFogFiniteandInfinite}, we present the confidence level under different switching costs for both fog and cloud latency with and without AdaptiveFog. \blu{The confidence of AdaptiveFog approaches that of a single MNO when the switching cost is large (e.g., switching to another MNO results in almost 50\% reduction in the confidence).} To compare the latency performance of different services, we present confidence levels under three maximum latency tolerable thresholds, 100 ms, 120 ms, and 150 ms. \blu{As the threshold value increases, the confidence degradation rate caused by the increase in switching cost becomes slower. This means that, when AdaptiveFog is applied, latency-tolerant applications become less sensitive to the switching cost, compared to latency-sensitive applications. In other words, AdaptiveFog is more suitable for applications that require very low service response time.} For both finite- and infinite-horizon optimizations, we observe that, when the switching cost is low, AdaptiveFog achieves almost 50\% improvement in confidence level for cloud latency, compared to the single MNO case. For fog latency, Figure \ref{Figure_ConfidenceVSSwitchingCostFogFiniteandInfinite} shows that AdaptiveFog achieves almost 30\% improvement in the confidence level for active road safety applications (e.g., 100 ms latency tolerance threshold). We also observe that in the finite-horizon scenario, the confidence level is more sensitive to the switching cost compared to the finite-horizon scenario.  Note that these simulation results are obtained using all of our driving traces to evaluate the performance improvement of AdaptiveFog. In some specific locations such as those when two MNOs exhibit high variations in latency performance, eNB deployment density, etc., the performance improvement achieved by AdaptiveFog is higher.

\blu{
In Figures \ref{Figure_ConfidenceVSSwitchingCostFogFiniteandInfiniteMyopic} and \ref{Figure_ConfidenceVSSwitchingCostCloudFiniteandInfiniteMyopic}, we compare the latency confidence level under AdaptiveFog with the myopic strategy (i.e., the UE tries to maximize the instantaneous utility by switching to different MNOs at each time slot) under different switching costs. As mentioned earlier, if the switching cost is small, AdaptiveFog achieves the same confidence level as the myopic strategy. As the switching cost increases, the performance difference between the two strategies will first increase and then converge to zero at high switching costs. In other words, the UE has no incentive to switch between networks under either strategy when the switching cost is high. We can also observe that, generally speaking, the gap in the fog latency confidence level between the two strategies is smaller than that for the cloud latency. This is because cloud latencies experienced in the two MNO networks exhibit more temporal and spatial variations compared to fog latencies. Therefore, if network adaptation can be carefully optimized to maximize the long-term performance, the benefit becomes more noticeable for the overall cloud latency confidence. Note that both AdaptiveFog and myopic strategies require information such as the K-R distance and switching cost at all the driving locations, and the decision thresholds must be pre-calculated and pre-stored at the UE or fog node. In Figures \ref{Figure_ConfidenceVSSwitchingCostFogFiniteandInfiniteMyopic} and \ref{Figure_ConfidenceVSSwitchingCostCloudFiniteandInfiniteMyopic}, we only consider a half week of driving data for training and performance evaluation with a limited driving time (around 2 hours per day). The performance improvement offered by AdaptiveFog is expected to become larger when the training and evaluation time is longer.
}



%
%



\section{Conclusions}
This paper reported a city-wide measurement campaign of the wireless access latency between vehicles and a fog computing system that is connected through commercially available LTE networks. 
We observed that the latency performance of different LTE networks can exhibit significant spatial variations and no MNO offers consistently better performance than the other.
A  novel networking and server adaptation framework, referred to as AdaptiveFog, was proposed, which allows vehicles to autonomously and dynamically connect to different LTE networks and fog or cloud servers. 
The main objective of AdaptiveFog is to maximize the confidence levels of various supported services with minimal switching between LTE networks.
An empirical spatial statistic model was established to characterize the spatial variations in latency across various locations of the city. We introduced the weighted K-R distance to quantify the performance gap between different LTE networks. 
A simple threshold-based policy was derived for a moving vehicle to sequentially switch to the optimal MNO.  
Extensive simulations were performed. Our results show that AdaptiveFog achieves around 30\% to 50\% improvement in the confidence level for fog/cloud latencies. 

\section*{Acknowledgment}
Y. Xiao was supported in part by the National Natural Science Foundation of China under grant 62071193 and the Key R \& D Program of Hubei Province of China under grant 2020BAA002. M. Krunz was supported in part by NSF (grants CNS-1910348, CNS-1563655, CNS-1731164, CNS-1813401, and IIP-1822071) and by the Broadband Wireless Access \& Applications Center (BWAC). Any opinions, findings, conclusions, or recommendations expressed in this paper are those of the author(s) and do not necessarily reflect the views of NSF. The authors would like to thank Haris Volos from Denso International America for making the Delay Explorer app available.

\bibliography{reference}
\bibliographystyle{IEEEtran}

\begin{IEEEbiography}[{\includegraphics[width=1.1in,height=1.3in,clip,keepaspectratio]{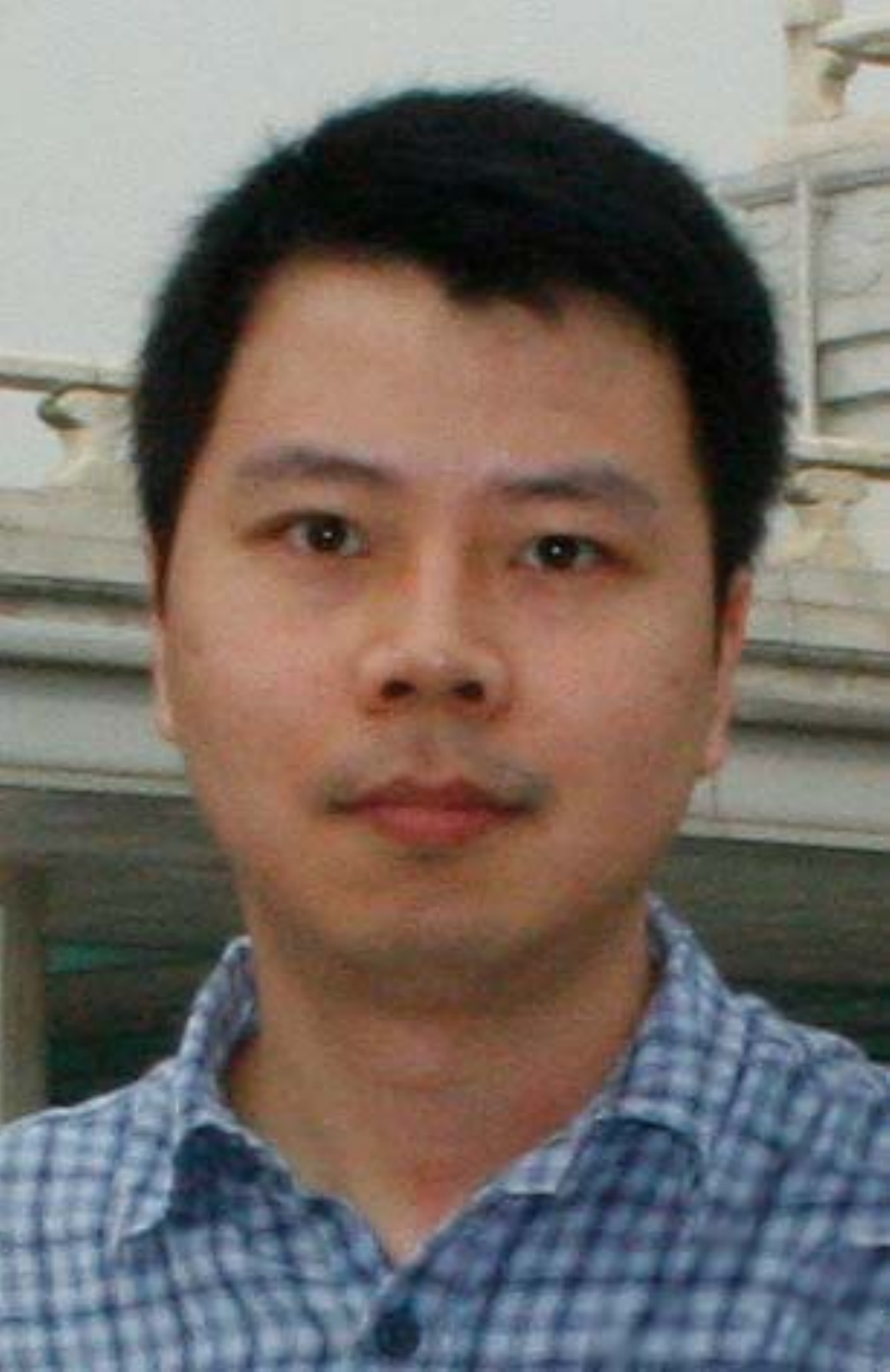}}]{Yong Xiao}(S'09-M'13-SM'15) received his B.S. degree in electrical engineering from China University of Geosciences, Wuhan, China in 2002, M.Sc. degree in telecommunication from Hong Kong University of Science and Technology in 2006, and his Ph. D degree in electrical and electronic engineering from Nanyang Technological University, Singapore in 2012. He is now a professor in the School of Electronic Information and Communications at the Huazhong University of Science and Technology (HUST), Wuhan, China. He is also the associate group leader of the network intelligence group of IMT-2030 (6G promoting group) and the vice director of 5G Verticals Innovation Laboratory at HUST. Before he joins HUST, he was a  research assistant professor in the Department of Electrical and Computer Engineering at the University of Arizona where he was also the center manager of the Broadband Wireless Access and Applications Center (BWAC), an NSF Industry/University Cooperative Research Center (I/UCRC) led by the University of Arizona. His research interests include machine learning, game theory, distributed optimization, and their applications in cloud/fog/mobile edge computing, green communication systems, wireless communication networks, and Internet-of-Things (IoT).
%
\end{IEEEbiography}

\vskip -2\baselineskip plus -1fil


\begin{IEEEbiography}[{\includegraphics[width=1.1in,height=1.3in,clip,keepaspectratio]{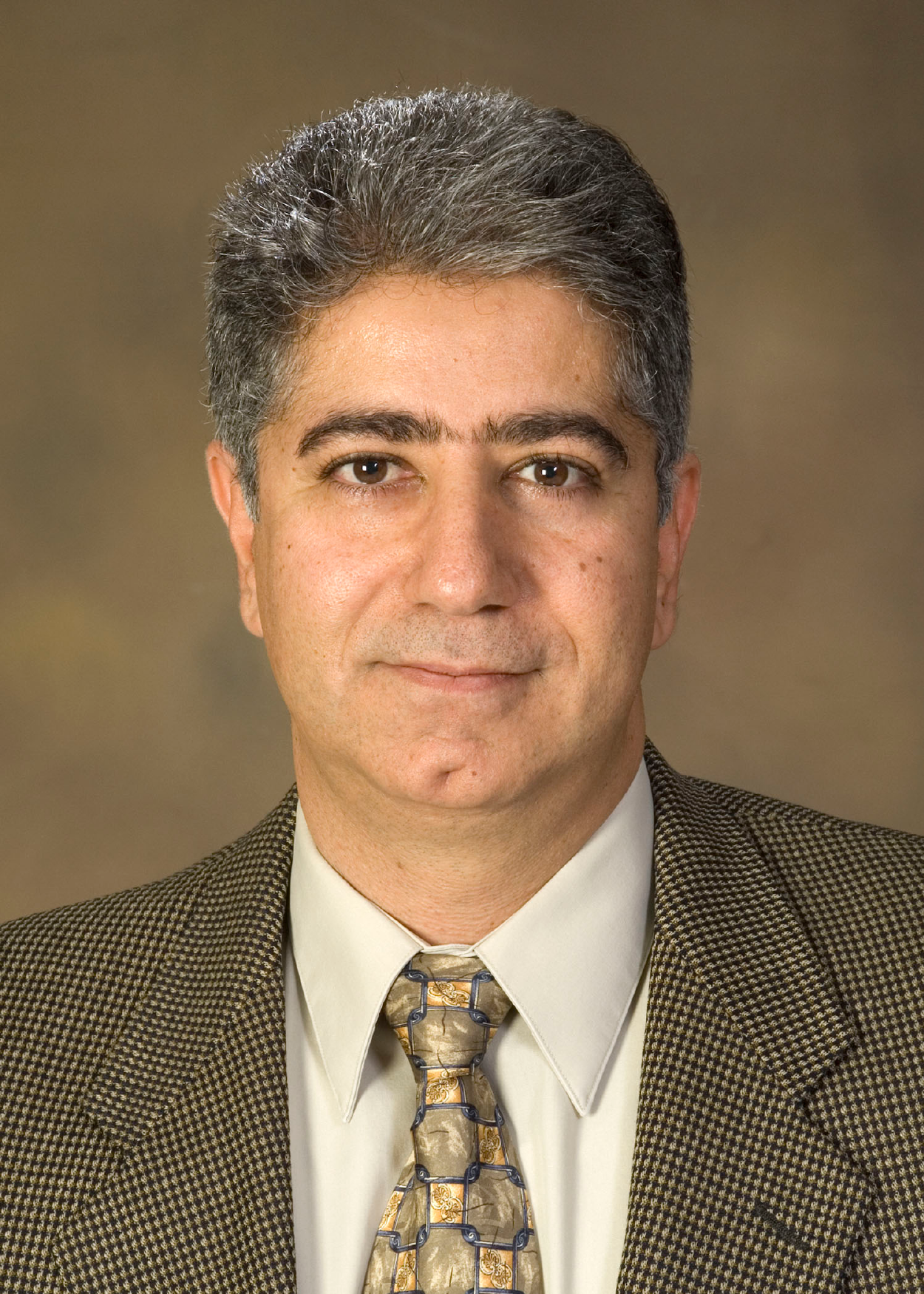}}]{Marwan Krunz}(S'93-M'95-SM'04-F'10) is the Kenneth VonBehren Endowed Professor in the Department of Electrical and Computer Engineering at the University of Arizona (UA). He is also the director of the Broadband Wireless Access and Applications Center (BWAC), a joint NSF/industry consortium that includes UA (lead site), North Carolina State University, University of Mississippi, and Catholic University of America, as well as 18+ members from industry and national labs. Dr. Krunz received the Ph.D. degree in electrical engineering from Michigan State University in July 1995. He joined the University of Arizona in January 1997, after a brief postdoctoral stint at the University of Maryland, College Park. In 2010, he was a Visiting Chair of Excellence (``Catedra de Excelencia") at the University of Carlos III de Madrid (Spain). He held numerous other short-term research positions at the University Technology Sydney, Australia (2016), University of Paris V, INRIA-Sophia Antipolis, France , University of Paris VI, HP Labs, Palo Alto, and US West Advanced Technologies. He was the Editor-in-Chief for the IEEE Transactions on Mobile Computing, and he previously served on the editorial boards of numerous journals, including IEEE/ACM Trans. on Networking, IEEE Trans. on Cognitive Communications and Networking, and others.
\end{IEEEbiography}
\end{document}